\documentclass[fleqn,usenatbib]{mnras}

\usepackage[T1]{fontenc}

\DeclareRobustCommand{\VAN}[3]{#2}
\let\VANthebibliography\thebibliography
\def\thebibliography{\DeclareRobustCommand{\VAN}[3]{##3}\VANthebibliography}

\usepackage{graphicx}	
\usepackage{amsmath}	
\usepackage{amssymb}	
\usepackage{float}
\usepackage{diagbox}

\usepackage{etoolbox}
\makeatletter
\patchcmd\@combinedblfloats{\box\@outputbox}{\unvbox\@outputbox}{}
\makeatother

\usepackage{multicol}        
\usepackage{bm}		
\usepackage{pdflscape}	
\usepackage{subfig}

\def\msun{\ifmmode \mathrm{M}_{\odot} \else M$_{\odot}$\fi}

\usepackage{newtxtext,newtxmath}

\title[Lensed star forming clumps at $2\lesssim z \lesssim6$]{Exploring the physical properties of lensed star-forming clumps at $2\lesssim z \lesssim6$}

\author[U. Me\v{s}tri\'{c} et al.]{
U. Me\v{s}tri\'{c},$^{1}$\thanks{E-mail: uros.mestric@inaf.it}
E. Vanzella,$^{1}$
A. Zanella,$^{2}$
M. Castellano,$^{3}$
F. Calura,$^{1}$
P. Rosati,$^{4,1}$
P. Bergamini,$^{5}$\newauthor
A. Mercurio,$^{6}$ 
M. Meneghetti,$^{1}$
C. Grillo,$^{5}$
G.B. Caminha,$^{7}$ 
M. Nonino,$^{8}$ 
E. Merlin,$^{3}$ 
G. Cupani,$^{8}$ 
E. Sani$^{9}$
\\
\\
$^{1}$INAF -- OAS, Osservatorio di Astrofisica e Scienza dello Spazio di Bologna, via Gobetti 93/3, I-40129 Bologna, Italy\\
$^{2}$INAF -- Osservatorio Astronomico di Padova, Vicolo Osservatorio 5, 35122, Padova, Italy \label{inafpd}\\
$^{3}$INAF -- Osservatorio Astronomico di Roma, Via Frascati 33, I-00078 Monte Porzio Catone (RM), Italy\\
$^{4}$Dipartimento di Fisica e Scienze della Terra, Universit\`a degli Studi di Ferrara, via Saragat 1, I-44122 Ferrara, Italy \label{unife}\\
$^{5}$Dipartimento di Fisica, Universit\`a  degli Studi di Milano, via Celoria 16, I-20133 Milano, Italy \label{unimi} \\
$^{6}$INAF -- Osservatorio Astronomico di Capodimonte, Via Moiariello 16, I-80131 Napoli, Italy \label{inafna}\\
$^{7}$Max-Planck-Institut f\"ur Astrophysik, Karl-Schwarzschild-Str. 1, D-85748 Garching, Germany\\
$^{8}$INAF -- Osservatorio Astronomico di Trieste, via G. B. Tiepolo 11, I-34143, Trieste, Italy \label{inafts}\\
$^{9}$European Southern Observatory, Alonso de Cordova 3107, Casilla 19, Santiago 19001, Chile \label{ESO}\\
}

\date{Accepted XXX. Received YYY; in original form ZZZ}

\pubyear{2015}

\begin{document}
\label{firstpage}
\pagerange{\pageref{firstpage}--\pageref{lastpage}}
\maketitle

\begin{abstract}
We study the physical properties (size, stellar mass, luminosity, star formation rate) and scaling relations for a sample of 166 star-forming clumps with redshift $z \sim 2-6.2$. They are magnified by the Hubble Frontier Field galaxy cluster MACS~J0416 and have robust lensing magnification ($2\lesssim \mu \lesssim 82$) computed by using our high-precision lens model, based on 182 multiple images. Our sample extends by $\sim 3$ times the number of spectroscopically-confirmed lensed clumps at $z \gtrsim 2$.  We identify clumps in ultraviolet continuum images and find that, whenever the effective spatial resolution (enhanced by gravitational lensing) increases, they fragment into smaller entities, likely reflecting the hierarchically-organized nature of star formation.
Kpc-scale clumps, most commonly observed in field, are not found in our sample. 
The physical properties of our sample extend the parameter space typically probed by $z \gtrsim 1$ field observations and simulations, by populating the low mass (M$_\star \lesssim 10^7$ M$_\odot$), low star formation rate (SFR $\lesssim 0.5$ M$_\odot$ yr$^{-1}$), and small size (R$_\mathrm{eff} \lesssim 100$ pc) regime.
The new domain probed by our study approaches the regime of compact stellar complexes and star clusters. 
In the mass-size plane, our sample spans the region between galaxies and globular clusters, with a few clumps in the region populated by young star clusters and globular-clusters.
For the bulk of our sample, we measure star-formation rates which are higher than  those observed locally in compact stellar systems, indicating different conditions for star formation at high redshift than in the local Universe.

\end{abstract}




\section{Introduction}
Investigating galaxies at sub-kpc scales is an essential step to understand the mechanisms that drive galaxy formation and evolution in the early Universe.
Currently, resolving structures at sub-kpc scales at high redshift is hardly achievable with deep ground-based adaptive optics (AO) or space-based observations.
Conversely, the combination of deep, high-resolution imaging and spectroscopic observations performed on gravitationally lensed targets enhances the resolving power and enables us to reveal sub-kpc structures and their physical properties at cosmological distances \citep[e.g.,][]{Livermore2012, Adamo2013ApJ, Livermore2015, Vanzella2016, Vanzella2017MNRAS, Rigby2017, Dessauges-Zavadsky2017, Johnson2017, Cava2018,  mirka19, Rivera2019, Vanzell2019MNRAS_clumps, Vanzella2021A&A, Bouwens2021, Iani2021MNRAS}.

In the last few decades, different studies revealed that high-redshift star-forming galaxies (SFGs, $z\sim1-3$) have a clumpy morphology, which differs from the spiral-like structure of most star-forming  galaxies in the local Universe \citep[e.g][]{Cowie1995,Elmegreen2005, Elmegreen2007, Conselice2014}.
High-redshift clumps have masses $10^{7} - 10^{9}\msun$, sizes $\sim0.1 - 1\rm kpc$, and usually they are characterized by a high star-formation rate \citep[e.g., ][]{Elemgreen2009, Jones2010, Genzel2011, Guo2012, Murata2014, Zanella2015, Shibuya2015, Guo2018, Zanella2019}.
Currently, the clumpy morphology of high-redshift galaxies has been explained with two formation scenarios: 
a) \textit{in situ}, the galaxy gas-rich disc fragments due to gravitational instability giving rise to massive star-forming regions \citep[e.g.,][]{Noguchi1999, Bournaud2008, Dekel2009_2, Hinojosa-Goni2016} or,
b) \textit{ex situ}, galaxies undergo a merger process during which the satellite is stripped and its nucleus appears as a massive clump \citep[e.g][]{Somerville2000,Hopkins2008, Puech2010, Straughn2015, Ribeiro2017}.

However, it is still unclear what is the relative contribution of these two channels to clumpy star formation.
The number of galaxies with clumpy morphology is higher than the expected merger rate among systems at high redshift, which makes it difficult to justify the formation of clumps through merging scenario alone \citep[e.g][]{Dekel2009_2, Hopkins2010}. 
Furthermore, some clumps with young ages ($\lesssim10\rm Myr$) have been reported \citep{Schreiber2011, Zanella2015}, as expected for clumps originating from disk fragmentation \citep{Bournaud2016}.  
Observational studies targeting $z\sim1 - 3$ galaxies suggests that $\sim 70$\% of clumps have an \textit{in-situ} origin, whereas the remaining ones have older ages and larger sizes and masses, consistent with being remnants of stripped satellites \citep{Zanella2019}.

Also the fate of clumps is still unclear and the results from simulations are highly dependent on the adopted assumptions (e.g., initial gas fraction of the disks, feedback recipes, \citealt{Fensch2021}). 
Hydrodynamical simulations presented in e.g. \cite{Murray2010, Genel2012, Tamburello2015, Oklopcic2017, Buck2017} suggest that massive clumps have a lifetime of $\sim20\,\rm Myr$, as stellar feedback quickly disrupts them, while less massive clumps tend to merge with more massive ones.
Conversely, \cite{Bournaud2014, Ceverino2010, Mandelker2014, Mandelker2017} predict that massive clumps can survive a few hundred Myr, until they merge with the galaxy bulge.

Simulations predict that the kpc-size clumps, which are typically observed in blank fields, have substructures which are blended and remains undetected due to the lack of resolution \citep[e.g][]{Meng2020MNRAS, Behrendt2016, Tamburello2015}. 
Indeed, recent observations of strongly lensed systems at $z\sim 1-3$ reveal the existence of clumps with sizes of $\sim100\,\rm pc$ \citep{Livermore2012, Livermore2015, Cava2018}, down to $\sim\! 10\,\rm pc$ \citep{Rigby2017, Johnson2017} in some extremely magnified cases.
Our previous studies  \citep{Vanzella2017(D11), Vanzell2019MNRAS_clumps, Vanzella2020, Vanzella2021_sun, Calura2021} have further extended this work to spectroscopically confirmed compact stellar structures, with sizes down to $\lesssim10\,\rm pc$, which are interpreted as young massive clusters and/or globular cluster precursors at $z\sim3-6$ (see also \citealt{Bouwens2021}). 
This demonstrates the importance of gravitational lensing to explore compact  stellar structures considered as building blocks of high-redshift star-forming galaxies.

Global relations among different physical properties (e.g., effective radius, stellar mass, star-formation rate (SFR)), as well as their redshift evolution, are widely used to understand the nature and evolution of galaxies and smaller stellar systems \citep[i.e][]{Kormendy1977,Bender1992, Bernardi2003, Kormendy2009, Misgeld&Hilker2011, Norris2014, Shibuya2015}.

In this work, we analyze the physical properties and scaling relations of clumps hosted by galaxies $z \sim 2 - 6$, which are gravitationally lensed by the Hubble Frontier Field (HFF) galaxy cluster MACS~J0416.1-2403 (hereafter MACS~J0416).
The paper is organized as follows:
Section \ref{Data} describes the data used in this work and the methodology used to select lensed clumps. Section \ref{subsec:clump_properties} explains how structural and photometric parameters are measured and physical properties of the clumps are inferred. 
Section \ref{comparison_with_sim} compares the physical properties of the MACS~J0416 sample with results from previous observations and simulations. 
In Section \ref{scaling relations}, we present the mass - SFR, size - SFR and mass - stellar surface density relations for our clumps sample. 
Finally in Section \ref{Summary}, we summarize our results and conclusions. Throughout the paper we adopt a flat $\Lambda$CDM cosmology with $\Omega_\mathrm{m} = 0.3$, $\Omega_\Lambda = 0.7$,
and $\rm H_{0}=70\, \rm km\,\rm s^{-1} \rm Mpc^{-1}$. All magnitudes are AB magnitudes \citep{Oke1974}, we adopt a \cite{Salpeter1955} stellar initial mass function and all measured physical parameters for our sample are corrected for lensing magnification unless differently stated.

\section{Data set and sample selection} \label{Data}

\subsection{Data and lens model}
\label{subsec:dataset}

Deep, multi-wavelength observations are publicly available in the MACS~J0416 field. 
This work is mainly based on \textit{HST}/ACS F814W and \textit{HST}/WFC3 F105W 30mas images \citep{Lotz2017,Koekemoer2014} used for measuring structural parameters and ASTRODEEP PSF matched images \textit{HST}/ACS F435W, F606W, F814W and \textit{HST}/WFC3 F105W, F125W, F140W, F160W \citep{Merlin2016A&A}, used for photometry and VLT-HAWKI/Ks imaging \citep{Brammer2016}.

Excellent spectroscopic coverage of this field, which is a critical component of this work since it affects the accuracy and precision of the lens model,  is based on deep VLT/MUSE integral field spectroscopic observations.
Details on the MUSE observations and data reduction are described in \citealt{Vanzella2021A&A} (V21). Building on and extending the previous study of  \citealt{Caminha_macs0416, richard2021}, V21 also presented redshift measurements by identifying  48 background galaxies at $0.9 < z < 6.2$, which are lensed into 182 bona-fide multiple images. 
The positions of these lensed sources, all spectroscopically confirmed, were used to constrain the strong lensing model in this work, which is described in  \citealt{Bergamini2021} (B21). 
Such an unprecedented large number of bona-fide multiple images, combined with a highly complete sample of cluster galaxies (80\% of which are spectrosopically confirmed) and MUSE-based measurements of the internal velocity dispersion of 64 cluster galaxies (down to 5 magnitudes fainter than the BCG), led to an accurate reconstruction of the mass distribution of MACS~J0416. 
The lens model is characterized by an r.m.s. displacement between the 182 observed and model-predicted image positions of only 0.40\arcsec.
The accurate magnification values which can be derived with the B21 lens model, with quantifiable statistical errors in different magnification regimes, particularly in the vicinity of critical lines (see B21), are used in this work to infer robust intrinsic properties of high-z lensed sources.

The average magnification ($\mu_{\rm avg}$) of each clump presented in this work is evaluated from magnification maps, derived with the B21 lens model.
The $\mu_{\rm avg}$ is calculated using a small aperture of $2\times2$ pixels (since our sources are compact), where the evaluated $\mu_{\rm avg}$ is compatible with the magnification of a point-like source.

\subsection{Clumps identification}
\label{subsec:identification}
We applied the following procedure to identify clumps and deblend their emission from that of the underlying diffuse host galaxy, as well as the contamination from nearby bright objects and intracluster light.
We considered cutouts of $15''\times 15''$, centered on each of the sources reported in the catalogue by \cite{Vanzella2021A&A}. 
With the {\tt IRAF} \citep{Tody1986} task {\tt MEDIAN}, we smooth each image with a boxcar filter. 
We subtract the smooth image from the original cutout to obtain a contrast enhanced image. We then use these new cutouts to detect clumps both visually and automatically with {\tt SExtractor (v2.24)} \citep[SE,][]{Bertin1996A&A}. 
During this process, it is key to select the appropriate smoothing box size, as the host light will be over-subtracted or under-subtracted if the box is too small or too large, respectively.
For our purposes, a good compromise was achieved with box sizes $\sim2-3$ times larger than each source. 
In most cases, the smoothing box size was 21 -- 31 pixels\footnote{An example of such a light removal procedure is shown in Figure~\ref{galfit}~(e).}. 
Such a procedure has been used in the literature to study clumps in field galaxies \citep{Guo2015} as well as to remove the contamination of the host galaxy halo before modelling ultra-compact dwarfs \citep{Norris2011}.

\subsection{Final clumps sample}
\label{subsec:final_sample}
Our final sample consists of 67 systems (48 of them showing multiple images), and in total 166 clumps have been detected. 
We define a "clump" as any compact source detected in the contrast images described in Section \ref{subsec:identification}.
All the clumps in our sample have spectroscopic redshifts based on MUSE deep observations \citep[][]{Caminha_macs0416, Vanzella2021A&A, richard2021} and spans the redshift range $z\sim2-6$.

\begin{figure}
  \centering
  \subfloat{\includegraphics[width=\columnwidth]{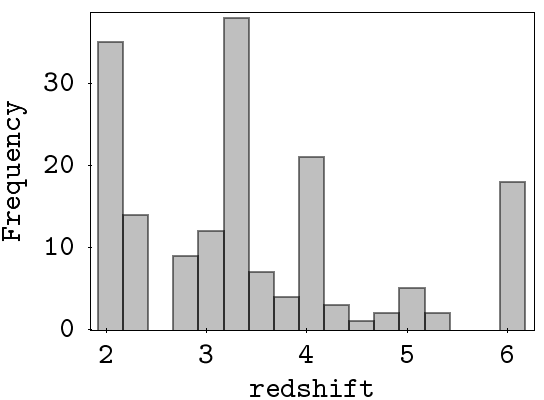}} \\
  \subfloat{\includegraphics[width=\columnwidth]{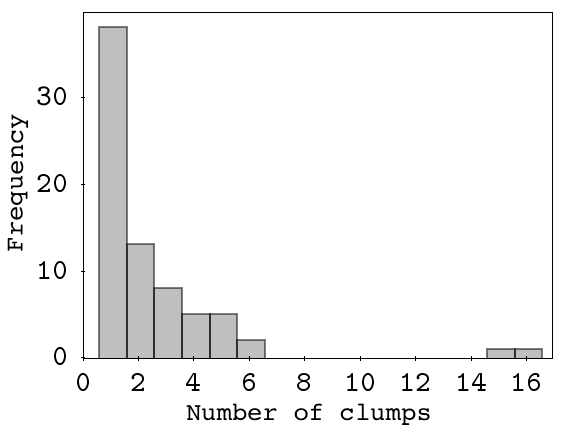}}
  \caption{The MACS~J0416 sample of lensed sources. \textbf{Top panel}: redshift distribution of the unique clumps (no multiple images are reported).  \textbf{Bottom panel}: distribution of the number of clumps per system.} 
  \label{fig:z_distr}
\end{figure}

In Figure \ref{fig:z_distr}, we show the redshift distribution of all clumps and the number of detected clumps per system.
The peaks visible in Figure~\ref{fig:z_distr} (upper panel) at $z\sim2, 3.5$ and 6 are the combination of two effects: (1) multiple clumps have the same redshift because hosted in the same galaxy; (2) there are galaxies at the same redshift possibly belonging to a physical larger structure. 
However, the limited sky coverage currently provided by MUSE ($1'\times1'$ FoV) in combination with a reduced effective area probed in the gravitational lens field (lensing cross section) is not sufficient to characterize possible Mpc-scale structure associated to the selected clumps.

\begin{figure}
  \centering
  \includegraphics[width=\columnwidth]{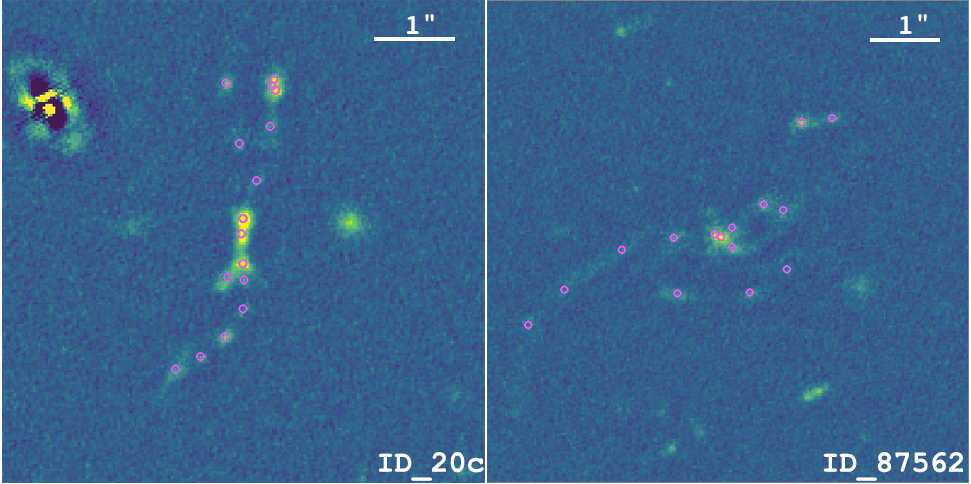}
  \caption{The two systems with the largest number of detected clumps at redshift $z=3.219$ (ID20c, left) and $z=2.089$ (ID87562, right). Small magenta circles indicate individual clumps. System ID20c has on average $\mu_{\rm avg} \approx 5.5$ and 16 detected clumps, while system ID87562 has a on average $\mu_{\rm avg} \approx 3.1$ and 15 detected clumps.}
  \label{fig:2systems}
\end{figure}

We find in most of the cases 1--2 clumps per system as shown in Figure \ref{fig:z_distr} (bottom panel).
In our sample there are also two systems hosting 15 and 16 clumps. 
The system ID20c with the highest number of detected clumps (16), at $z=3.219$, has an irregular shape, with five clumps located in the central region (Figure \ref{fig:2systems} left).
While the other system, ID87562, is a face-on spiral galaxy  with 15 detected clumps at redshift $z=2.089$.
The majority of the clumps are presumably star-forming regions located in the spiral arms of the hosting galaxy (Figure \ref{fig:2systems} right).

\section{Measuring clump properties} \label{subsec:clump_properties}

\subsection{Clump structural parameters}
\label{subsec:structural_parameters}
To measure the structural properties of clumps, we use the \textit{HST} F814W and F105W bands, as they offer the best compromise in terms of spatial resolution (FWHM PSF $\sim 0.13"$(F814W) and $\sim 0.15"$(F105W), pixel scale $\sim$ 0.03 arcsec/pixel) and depth (limiting AB magnitude $\sim 29$). 
Such images probe rest-frame $\lambda \simeq 1400-2000$\AA\ at $2<z<6.3$. 
In particular, we use the F105W-band for sources at $z\gtrsim5.1$, where the Ly$\alpha$-break ($\lambda<1216$\AA) enters the F814W filter.

\begin{figure*}
  \centering
  \includegraphics[width=\textwidth]{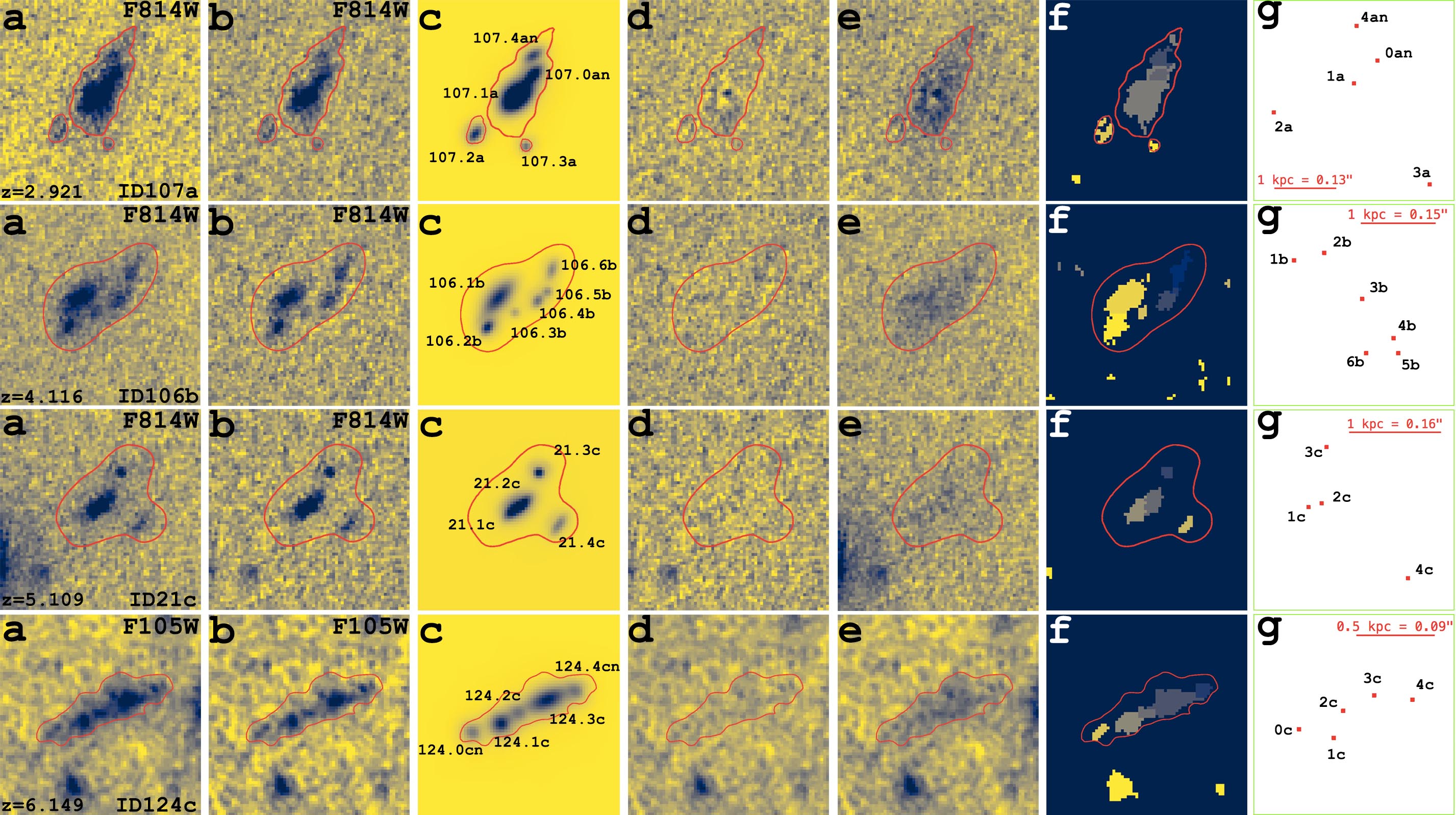}
  \caption{An example of {\tt GALFIT} modeling and host light removal applied to four typical sources from our sample at $z\approx 3-6$: system ID107a with $\mu_{\rm avg}=5.75$, system ID106b with $\mu_{\rm avg}=11.3$, system ID21 with $\mu_{\rm avg}=11.4$ and system ID124c with $\mu_{\rm avg}=12$.   \textbf{Panel a}: the original \textit{HST} F814W (F105W for system ID124c) image. \textbf{Panel b}: \textit{HST} F814W (F105W for system ID124c) image after diffuse light subtraction. \textbf{Panel c}: {\tt GALFIT} model. \textbf{Panel d}: residual map obtained by subtracting the model (panel c) from the host-subtracted image (panel b). \textbf{Panel e}: host galaxy light visible after subtracting the {\tt GALFIT} model (panel c) from the input \textit{HST} F814W image (panel a). \textbf{Panel f}: the {\tt SExtractor}  segmentation map provided to {\tt GALFIT} together with the initial guesses of the structural parameters. All cutouts \textbf{a--f} on the image plane are 2.1\arcsec across. \textbf{Panel g}: shows the predicted positions of each clump and the physical scale of each system on the source plane (red crosses) using our lens model. The red contours mark the smooth surface brightness and shape of each system.}
  \label{galfit}
\end{figure*}

We run {\tt SExtractor} \citep[SE,][]{Bertin1996A&A} on the contrast enhanced images (as mentioned in Section \ref{subsec:identification}) to create a segmentation map, to identify the clumps and to obtain initial guesses of their structural parameters (e.g. coordinates, half-light radius, ISO magnitude, AUTO magnitude, circular aperture magnitude, as well as related uncertainties). 
We use these values as initial guesses for the two-dimensional modelling of the clump light profiles that we perform with the {\tt GALFIT (v3.0.7)} tool \citep{Peng2002AJ, Peng_2010}. 
When multiple clumps in a single galaxy are present, we model them one at a time, using the segmentation to unmask the clump of interest.
When clumps are faint and/or positioned near other bright sources, {\tt SExtractor} is not converging to a reasonable segmentation image and therefore we provide manually the initial parameters for the {\tt GALFIT} modelling, after visual inspection. This happened for $\sim\! 15$ clumps ($\sim$9\% of the sample).
Such a procedure produces good results and convergence. We consider a fit as successful when the residual two-dimensional map (obtained by subtracting the model from the input image) is smooth (i.e., no structures are detectable) and the normalized residuals (obtained by normalizing the residuals map by the input image) are smaller than 20\% on average.  An example of {\tt SExtractor} segmentation and {\tt GALFIT} modeling is shown in Figure \ref{galfit}, where
we also show (in panel g) the predicted positions of each clump on the source plane obtained from our lens model. 
By converting angular separation among the clumps into physical distances, we note that some clumps are less than 100pc apart, while their parent systems extend over 1-2 kpc.

\begin{figure}
  \centering
  \subfloat{\includegraphics[width=\columnwidth]{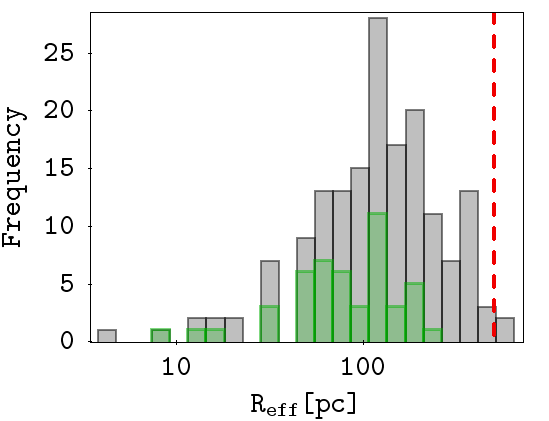}} \\
  \subfloat{\includegraphics[width=\columnwidth]{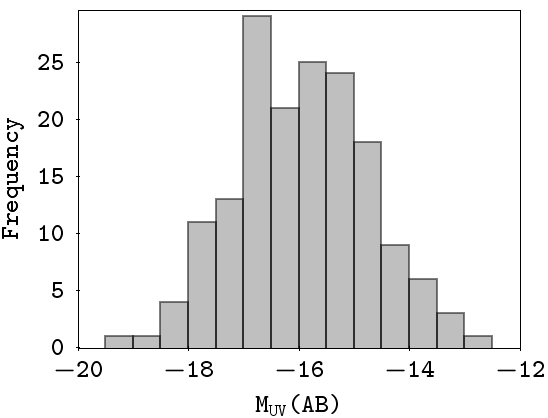}}
  \caption{Distribution of structural properties of our sample. \textbf{Top panel:} distribution of measured effective radius of the clumps. The red vertical dashed line denotes the smallest clump size detected in field and the green histogram shows the distribution of clumps with estimated sizes as upper limits. \textbf{Bottom panel}: distribution of absolute UV-rest frame magnitudes for the presented sample.} 
  \label{fig:reff_and_Muv}
\end{figure}

To model the surface brightness of the clumps, we consider S\'ersic profiles with indexes $n=0.5$ (Gaussian profile) or 1 (exponential profile)\footnote{Higher values of $n$ do not change the resulting solution significantly and leaving $n$ as a free parameter for faint knots typically weaken the convergence of the process}.
Either index value is used based on the quality of residual image produced by {\tt GALFIT}.
In 28 cases {\tt GALFIT} modelling could not converge and we used a PSF profile instead. This happened for faint, compact clumps.
In the cases where the PSF model was used, we set an upper limit on the effective radius equal to the half-width-half-maximum of the PSF. 
Finally, when the Gaussian or exponential model provide a solution with $\rm R_{eff}<1$ pixel, the clump effective radius is considered as an upper limit and set to 1 pixel (corresponding to 30 mas).
The distribution of effective radii and rest-frame absolute UV magnitudes (M$_{\rm UV}$) for the entire sample are presented in Figure~\ref{fig:reff_and_Muv}.

\subsection{Photometric measurements}
\label{subsec:photometric_measurements}

Photometric measurements on confirmed clumps were performed using the data reduced by the {\tt ASTRODEEP} collaboration \citep{Merlin2016A&A, Castellano2016A&A}, where the intracluster light and the brightest cluster members are subtracted and all images are PSF-matched, and then the photometric tool A-PHOT \citep{Merlin2019APHOT} is utilized.
Specifically, first, we identify the subsample of clumps in common with the {\rm ASTRODEEP} published catalog \citep[][]{Merlin2016A&A, Castellano2016A&A}, in particular those showing a clean detection, not contaminated by nearby sources (37 objects). 
We refer to this sample as {\it category 1} and rely on the public photometric catalog for what concerns their SEDs. 
For this category, SEDs have multi-band coverage including VLT-HAWKI/Ks and IRAC channels 1 and 2. 
The remaining objects in our sample are not present in the {\rm ASTRODEEP} catalog, because they are too faint 
or not individually detected on larger images. 
For these cases we perform our own photometric extraction, including the HAWKI/Ks band (excluding the IRAC photometry, whose PSF is too large). 
For the \textit{HST} photometric measurements, we use previously PSF-matched images \citep{Merlin2016A&A} observed in seven \textit{HST} broadband filters F435W, F606W, F814W, F105W, F125W, F140W, F160W.
We perform circular aperture photometry for all clumps of the sample with an aperture of $0.27''$ in diameter, which is slightly larger than the PSF FWHM  of PSF-matched images ($0.2''$) but small enough to still capture the clumps analysed in this work. 
In this way the colors are preserved. Subsequently, each SED is  aperture-corrected (AC) by rescaling it to the total {\rm GALFIT} magnitude in the same reference band. 
AC values spans the interval $1-4$, depending on the compactness of the object (the more compact the smaller the correction).

To include the HAWKI/Ks band, the sample is further split into two categories: a sample of clumps for which no A-PHOT signal is detected in the Ks band within an aperture of $0.4''$ diameter, SNR$<$5 (dubbed as {\it category \rm 2}, 78 objects) and the remaining subset with a Ks-band detection, SNR$>$5 (dubbed as {\it category \rm 3}, 51 objects). 
In case of category 2, $5\sigma$ upper limits are used for the Ks band, whereas for category 3 an estimate of the total Ks magnitude is obtained with {\rm GALFIT} after fixing all the structural parameters (except the magnitude) previously defined in the reference band for the same clump. 
This procedure is similar to the one adopted by \citet{Merlin2016A&A} with the T-PHOT tool,  \citep{T-PHOT2015,Merlin2016TPHOT} which uses analytical models as priors. 

Summarizing, category 1 benefits from the available {\rm ASTRODEEP} photometry. Categories 2 and 3 include those clumps not detected in {\rm ASTRODEEP}, which are respectively undetected or detected in the Ks band.

\subsection{Clump physical properties}
\label{subsec:Clump physical properties}

The physical properties (star-formation rate, mass, age) of the sample are derived by using the SED-fitting technique described in \cite{Castellano2016A&A}, however in our case we allow for ages as young as 1 Myr, extending the original {\rm ASTRODEEP} analysis.
The most relevant physical properties for this work are shown in Table \ref{tab:J0416data} (full version is available in electronic form).

In brief, the physical properties of the clumps  are computed by fitting \citet{Bruzual2003} (BC03) templates with the \verb|zphot.exe| code \citep[][]{Fontana2000} at the spectroscopic redshift. 
In the BC03 fit, we assume exponentially declining star-formation histories with e-folding time 0.1$\leq \tau \leq$ 15.0Myr, a \citet{Salpeter1955} initial mass function and we use both \citet{Calzetti2000} and Small Magellanic Cloud \citep{Prevot1984} extinction laws. 
Absorption by the intergalactic medium (IGM) is modeled following \citet{Fan2006}. 
We consider the following range of physical parameters: $0.0\leq E(B-V) \leq1.1$, Age $\geq 1$Myr (defined as the onset of the star-formation episode), metallicity $Z/Z_{\odot}=0.02,0.2,1.0,2.5$. 
We fit all the sources including the contribution from nebular continuum and line emission following \citet{Schaerer2009} under the assumption of an escape fraction of ionizing photons $f_{esc}=0.0$ \citep[see also][for details]{Castellano2014}.

\begin{figure*}
\begin{center}
    \subfloat{\includegraphics[width=.4\textwidth]{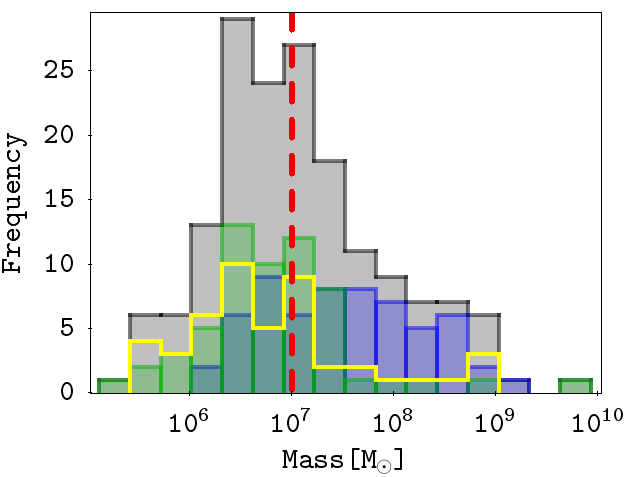}}
    \hspace{-0.01cm}\vspace{-0.05cm}
    \subfloat{\includegraphics[width=.4\textwidth]{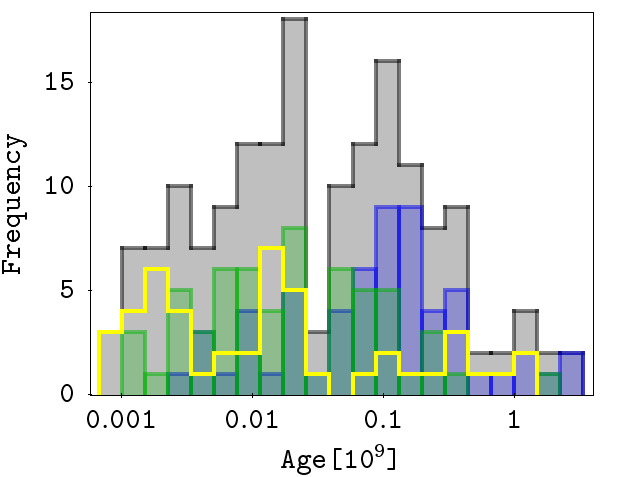}}
    \hspace{-0.01cm}\vspace{-0.05cm}

    \subfloat{\includegraphics[width=.4\textwidth]{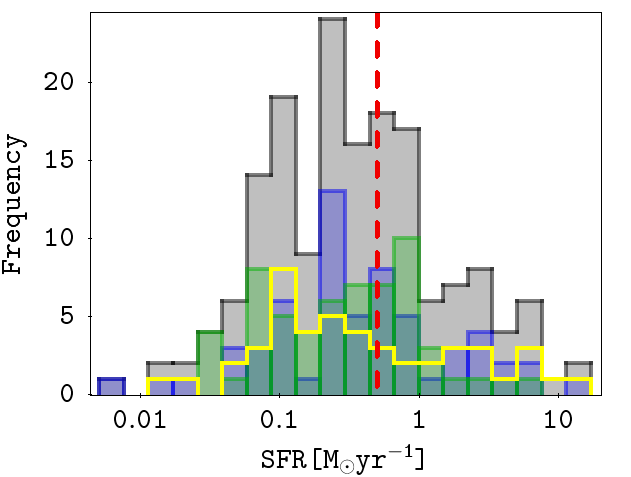}}
    \hspace{-0.01cm}\vspace{-0.05cm}
    \subfloat{\includegraphics[width=.4\textwidth]{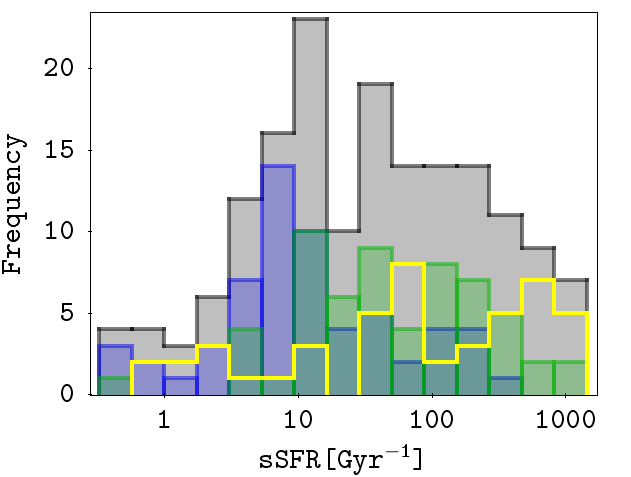}}
    \hspace{-0.01cm}\vspace{-0.05cm}

\caption{Physical properties of the MACS~J0416 sample derived by SED fitting. Grey histograms shows the entire sample, the blue ones represent clumps at $z=2-3$, the green clumps at $z=3-4$ and the yellow are clumps at $z=4-6.2$. The red vertical dashed line in the mass histogram indicates the smallest clump mass measured in field and red vertical dashed line in the SFR histogram indicates the typical SFR limit of clumps observed in field at $z>2$. Color coding in all four panels is the same.}
\label{fig:SED-results}
\end{center}
\end{figure*}

The physical properties derived by SED fitting are summarized in Figure \ref{fig:mass-size-mag-magnification} and \ref{fig:SED-results}. 
The clumps in our sample have stellar masses in the range $10^5 - 10^9$ M$_\odot$, star formation rates SFR $ \sim 0.001 - 10\, {\rm M}_\odot$ yr$^{-1}$, ages $\sim$ 1 Myr - 2.6 Gyr, and specific star formation rate sSFR $\sim 0.1 - 1000$ Gyr$^{-1}$. 
Clumps in the lowest redshift bin ($z \sim 2-3$) seem to be on average $\sim 0.4-0.7$dex more massive and $\sim0.6$ dex older than higher redshift clumps. They also have a $\sim 0.7$ dex lower sSFR. 
No clear evolution of SFR with redshift is observed (Figure \ref{fig:SED-results}).

\begin{table*}

\caption{Physical properties of the MACS~J0416 sample of star-forming clumps. }
\label{tab:J0416data}
\begin{tabular}{ccccccccccc}
\hline \hline
ID$^{1}$    & \begin{tabular}[c]{@{}c@{}}RA\\ {[}deg{]}\end{tabular} & \begin{tabular}[c]{@{}c@{}}DEC\\ {[}deg{]}\end{tabular} & $z_{\rm spec}$ & \begin{tabular}[c]{@{}c@{}}$\rm M_{\rm UV}$$^{2}$\\ {[}mag{]}\end{tabular} & \begin{tabular}[c]{@{}c@{}}Size$^3$\\ {[}pc{]}\end{tabular} & \begin{tabular}[c]{@{}c@{}}M$_\star$$^4$\\ {[}$\times 10^{8} \rm{ M}_\odot${]}\end{tabular} & \begin{tabular}[c]{@{}c@{}}SFR$^5$\\ {[}$\rm{M}_\odot yr^{-1}${]}\end{tabular} & $\mu_{\rm avg}$$^6$ & $\mu_{\rm tang}$$^7$ & Category$^8$ \\ \hline 
9.1a (1485)  & 64.04511  & -24.07235 & 3.290 & -17.7 & 161$\pm7$ & $0.2^{0.1}_{0.1}$    & $6.1^{7.1}_{4.2}$  & 4.2$\pm0.1$ & 2.8$\pm0.1$ & 1 \\
31.1c & 64.02341 & -24.07609 & 4.125 &-17.2 & 172$\pm9$    & $0.5^{0.8}_{0.2}$   & $0.3^{0.9}_{0.3}$  & 3.5$\pm0.1$ & 2.5$\pm0.1$ & 2 \\ 
36.1c & 64.02408 & -24.08090 & 1.963 &-17.1 & <55        & $2.0^{8.9}_{0.7}$ & $0.6^{2.5}_{0.4}$  & 6.0$\pm0.2$ & 4.6$\pm0.2$ & 3 \\
... & ... & ... & ...   & ...  & ...   & ...   & ... & ...  &  ... & ...     \\ \hline \hline

\end{tabular}

\begin{flushleft}
Notes: All physical parameters and uncertainties in the table are corrected for lensing magnification , adopting the magnification $\mu_{\rm avg}$ reported in column \#9 and $\mu_{\rm tang}$ in column \#10 for the Size (column \#6) . The full Table is available in electronic form.\\
$^{1}$ Clump ID, for clumps from \textit{category 1} ASTRODEEP ID is in the brackets.\\
$^{2}$ Absolute UV rest-frame magnitude.\\
$^{3}$ Half light radius, R$_{\rm eff}$.\\
$^{4}$ Stellar mass derived from SED fitting. Quoted uncertainties are 16 and 84 percentile range from SED fitting. \\
$^{5}$ Star formation rate derived from SED fitting. Quoted uncertainties are 16 and 84 percentile range from SED fitting. \\
$^{6}$ Average magnification \citep{Bergamini2021}.\\
$^{7}$ Tangential magnification \citep{Bergamini2021}.  \\
$^{8}$ Category of the clump (see the text for explanation).\\
\end{flushleft}

\end{table*}

\subsection{From observed to intrinsic properties and error propagation}
\label{subsec:From observed to intrinsic properties and error propagation}

In order to investigate the intrinsic properties of a lensed objects (in our case clumpy structures located in galaxies), it is crucial to have a robust cluster lens model and spectroscopic redshifts.
Only in this case we can properly correct for lensing magnification the measured properties (i.e. magnitudes, size, SFR, masses).

In this work, after measuring the half-light radius (R$_{\rm eff}$) for each clump in pixels (as described in \ref{subsec:structural_parameters}), the given values are translated to physical sizes (in parsec) and corrected for magnification. In this case, we use the tangential component of the magnification, $\mu_{\rm tang}$, since images near the tangential critical line are mostly elongated tangentially. 
On the other hand, the observed magnitudes, measured with {\tt GALFIT}, are corrected for the average magnification to obtain the intrinsic magnitudes $\rm m_{UV}$, and further converted to absolute magnitudes $\rm M_{UV}$ (rest frame).
Other physical properties (mass and SFR), discussed in this work are derived from SED fitting and their values and associated uncertainties are corrected for the average magnification.
The uncertainties on the magnification \citep[adopted from][]{Bergamini2021} and from SED fitting are added in quadrature to obtain the total uncertainties.

To evaluate the robustness of the measured sizes and magnitudes and their uncertaintieswe adopted an approach similar to the one described in \cite[APPENDIX A,][]{Vanzella2017(D11)}. 
We use the routines from the {\sc SkyLens} package \citep{Meneghetti2008,Meneghetti2017,Plazas2018} implemented in the new {\sc Forward Modeling} \citep[FM,][]{Bergamini2021,Meneghetti2021LNPbook} python code to create mock HST observations of stellar clumps of different sizes and intrinsic magnitudes. 
The simulated images also include a noise level (assumed to be Poisson-like) mimicking the real observations in $HST$ F105W band.

First, we test how robustly we can measure R$_{\rm eff}$ of lensed sources by running {\tt GALFIT} with a Monte Carlo approach. The range of intrinsic input magnitudes and sizes in the simulations is chosen to best describe the properties of the clumps in our sample: $24 \leq \rm m_{AB}\leq 30$ and $ 1 \leq \mathrm{R}_{\rm eff} \leq 7$ pixels. Table \ref{tab:J0416_uncertainties_Reff} summarizes the adopted magnitude-size pairs simulated with FM. 
The pairs marked with '$\times$' are cases which are not present in our sample.
We create 50 lensed mock images of the same source (at the position of clump 2b at z=6.14, where $\mu_{avg}=18.5$ and $\mu_{tang}=13.1$) varying the magnitude and size in the source plane as mentioned above, and fixing the Sersic index n=1. 
The emerging image in the lens (observed) plane is properly magnified by the FM procedure, in which random Poisson noise is also computed for each realization mimicking the real depth of the F105W image. 
{\tt GALFIT} has therefore been run 50 times for each magnitude-size pair leaving all parameters free to vary in the fitting procedure (X, Y positions, position angle -- PA, elipticity -- b/a, magnitude and n). 
The same has been done after fixing a few of them (X, Y positions, PA, b/a and n), especially in the faint domain (mag > 27).
Both approaches (leaving free all parameters or fixing some of them to the actual values) allow us to recover the input values (magnitude, sizes) with compatible medians and uncertainties in the bright magnitude regime (mag$<27$). 
For the faintest and most compact cases, as expected, the convergence is not always guaranteed and, as we did for the real clumps, it is useful to fix some of the input parameters.
Indeed, in the real sample of clumps, {\tt GALFIT} fitting has been performed in a semi-automatic way for the faint regime (mag > 27), after fixing n, PA and b/a quantities and visually inspecting each model and residual map produced by {\tt GALFIT}. 
Similarly, in the MC simulations we verified that when fixing Sersic to Gaussian (n=0.5) or exponential (n=1) for the faint and/or compact cases {\tt GALFIT} produces results consistent with the exponential case (n=1, the profile set in the source plane). Also the position angle is not relevant in the cases of circular symmetric shapes, which is frequent among the faint and compact clumps. 
For the elongated ones, such parameter is properly retrieved in the bright regime, whereas it is inferred manually and after visual inspection for the faint and extended ones.
We did not simulate such an user-dependent process, therefore for the faint regime of the MC simulations
we fix the position and Sersic to the right values and leave the rest free.
For each set of 50 MC {\tt GALFIT} runs we compute the median and the 16th and 84th percentiles of the resulting distributions of sizes and magnitudes. 
We report the obtained values in Table \ref{tab:J0416_uncertainties_Reff}.

As an additional test, we run {\tt GALFIT} manually on 51 simulated cases belonging to the faint regime (mag$>27-29.4$) and inserting radii in the range $1-4$ pixel without knowing the input parameters (performed blindly among the  authors) and applying the same methodology used for the real clumps. 
The relevant quantities, i.e. magnitude and effective radii, have been recovered with uncertainties consistent to those inferred from simulations: magnitudes are recovered with a global error of $\lesssim 0.2$ mag independently from the radius and magnitude, while the radius has been recovered with and error $\lesssim (30-50)\%$ in the range 27$<$mag$<$28, and $(50-100)\%$ at fainter magnitudes ($>28$).

\begin{table*}
\caption{Recovered R$_{\rm eff}$ in pixels with {\tt GALFIT }as a function of magnitude (mag).}
\label{tab:J0416_uncertainties_Reff}
\begin{tabular}{llllllll}
\hline \hline
\diagbox{$\rm mag$$^{1}$}{R$_{\rm eff}$[pix]$^{2}$}& 1  & 2  & 3  & 4  & 5  & 6  & 7  \\
\hline \hline
\\
 30 & $3.24^{+4.80}_{-1.87}$ & $\times$               & $\times$               & $\times$               & $\times$ & $\times$ & $\times$ \\
 29 & $1.46^{+1.93}_{-0.51}$ & $2.68^{+2.65}_{-1.31}$ & $\times$               & $\times$               & $\times$ & $\times$ & $\times$ \\
 28 & $1.19^{+0.57}_{-0.38}$ & $2.08^{+0.78}_{-0.66}$ & $2.84^{+1.38}_{-0.75}$ & $3.87^{+1.19}_{-1.21}$ & $\times$ & $\times$ & $\times$ \\
 27 & $1.04^{+0.37}_{-0.26}$ & $2.10^{+0.30}_{-0.45}$ & $2.94^{+0.46}_{-0.41}$ & $3.96^{+0.28}_{-0.65}$ & $4.79^{+0.49}_{-0.53}$ & $5.54^{+0.48}_{-0.37}$ & $6.35^{+1.02}_{-0.53}$ \\
 26 & $1.17^{+0.14}_{-0.22}$ & $2.16^{+0.12}_{-0.23}$ & $3.09^{+0.25}_{-0.27}$ & $3.96^{+0.17}_{-0.27}$ & $4.90^{+0.40}_{-0.25}$ & $5.74^{+0.30}_{-0.24}$ & $6.68^{+0.49}_{-0.33}$ \\
 25 & $1.19^{+0.08}_{-0.05}$ & $2.09^{+0.13}_{-0.06}$ & $3.05^{+0.09}_{-0.10}$ & $3.99^{+0.10}_{-0.12}$ & $4.96^{+0.10}_{-0.13}$ & $5.92^{+0.15}_{-0.21}$ & $6.75^{+0.23}_{-0.22}$ \\
 24 & $1.18^{+0.03}_{-0.04}$ & $2.22^{+0.08}_{-0.14}$ & $3.07^{+0.03}_{-0.03}$ & $4.0 ^{+0.06}_{-0.05}$ & $4.94^{+0.07}_{-0.04}$ & $5.91^{+0.07}_{-0.09}$ & $6.85^{+0.07}_{-0.16}$ \\
 \\ \hline \hline
\end{tabular}

\begin{flushleft}
Notes: Recovered R$_{\rm eff}$ are median values and lower and upper limits are 16th and 84th percentiles.\\
$^{1}$ Input observed magnitudes. \\
$^{2}$ Input half light radius in pixels\\
\end{flushleft}

\end{table*}

In a second step, we test the accuracy and precision of the magnitude measurements after removing the host light. 
To this goal, we add an additional light component (diffuse light) with R$_{\rm eff}=35$pix to the simulated clump. 
The amount of added diffuse light is equivalent to $\rm m_{AB}\sim 26.8(1), 26.1(2), 25.6(3)$, where numbers in brackets are telling how many times the host is brighter than the clump.
In all cases, we assume R$_{\rm eff}=3 \rm pix$ and $\rm m_{AB} = 26.8$ for the clump component.
For each host magnitude and given clump properties (R$_{\rm eff}=3 \rm pix$ and $\rm m_{AB} = 26.8$, n=1), we create 50 mock lensed images. 
For each of them, we remove the host light in the same way as we are doing for real images (described in \ref{subsec:identification}). Then, we use {\tt GALFIT} to recover the clump magnitude. 
As done earlier, we compute the median and the 16th and 84th percentiles of the measurement distributions for each clump magnitude (after host removal procedure) and report the results in Table \ref{tab:J0416_uncertainties_mag}. 
The results show the error on the measured clump magnitude of $\lesssim 0.2$ mag. 
Such uncertainty can affect applied AC correction and further mass and SFR estimates derived by SED fitting.
In other words, the host light-removing procedure in the most extreme cases can change the presented mass and SFR predictions by not more than $\sim 15\%$.
This is within the 16th and 84th percentile range given from SED fitting (reported uncertainties).

\begin{table}
\caption{Recovered magnitudes of the clump $(\rm mag_{\rm c})$ with $\rm mag_{\rm c}$=26.8 as a function of host magnitude $(\rm mag_{\rm h})$}
\label{tab:J0416_uncertainties_mag}
\begin{tabular}{llllllll}
\hline \hline
\diagbox{$\rm mag_{\rm c}$$^{1}$}{mag$_{\rm h}$$^{2}$}& 26.8  & 26.0  & 25.6 \\
\hline \hline
\\
 26.8 & $27.0^{+0.1}_{-0.1}$  & $26.9^{+0.1}_{-0.1}$ & $26.8^{+0.1}_{-0.5}$  \\
 \\ \hline \hline
\end{tabular}

\begin{flushleft}
Notes: Recovered magnitudes are median values and lower and upper limits are 16th and 84th percentiles.\\
$^{1}$ Clump input magnitudes.\\
$^{2}$ Magnitude of the host\\
\end{flushleft}

\end{table}

\section{Comparison with simulations}\label{comparison_with_sim}
 
\subsection{Simulations}
\label{subsec:sims}
We compare observational results with high-resolution hydrodynamic simulations of $z \gtrsim 2$ clumpy galaxies from the literature. In particular, we consider the following samples:
\begin{itemize}
    \item the simulations by \cite{Tamburello2017} have a resolution of 100 pc. They simulate  $\sim 15$ clumpy galaxies and in mock H$\alpha$ maps they identify clumps  with sizes R$_\mathrm{e} \sim 100 - 300$ pc and stellar mass M$_\star \sim 10^{7} - 10^{8}$ M$_\odot$;
    
    \item the simulation by \cite{Faure2021} is focused on clumpy galaxies at redshift $\sim 2$ which are gas-rich and disc-dominated.
    The stellar masses of the simulated galaxies are $3.1-10 \times 10^{10}$ M$_\odot$ and the resolution of the simulations are 500 and 12.2 pc per pixel.
    In their simulations, the identified clumps spans the stellar mass range $\sim 10^{6}-10^{10}\,\rm{M}_\odot$;
    
    \item the simulation by \cite{Bournaud2014} detailed evolution of giant clumps in four high redshift discs ($z=2$) is modeled.
    The resolution of the simulation is 3.5--7 pc with initial masses of the galaxies from $1.4-8.7 \times 10^{10}\,\rm{M}_\odot$.

\end{itemize}

\begin{figure*}
\begin{center}
    \subfloat{\includegraphics[width=.4\textwidth]{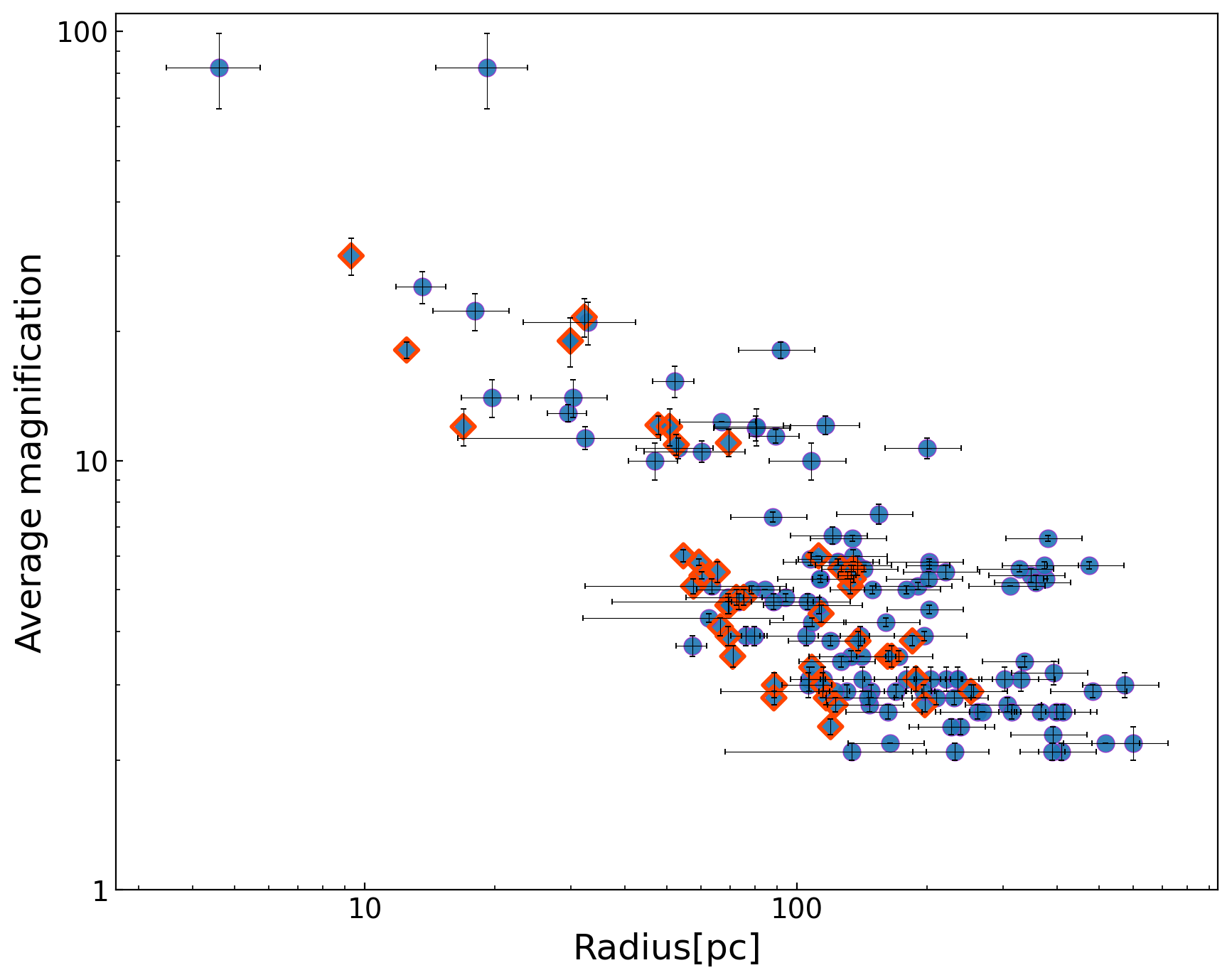}}
    \hspace{-0.01cm}\vspace{-0.05cm}
    \subfloat{\includegraphics[width=.4\textwidth]{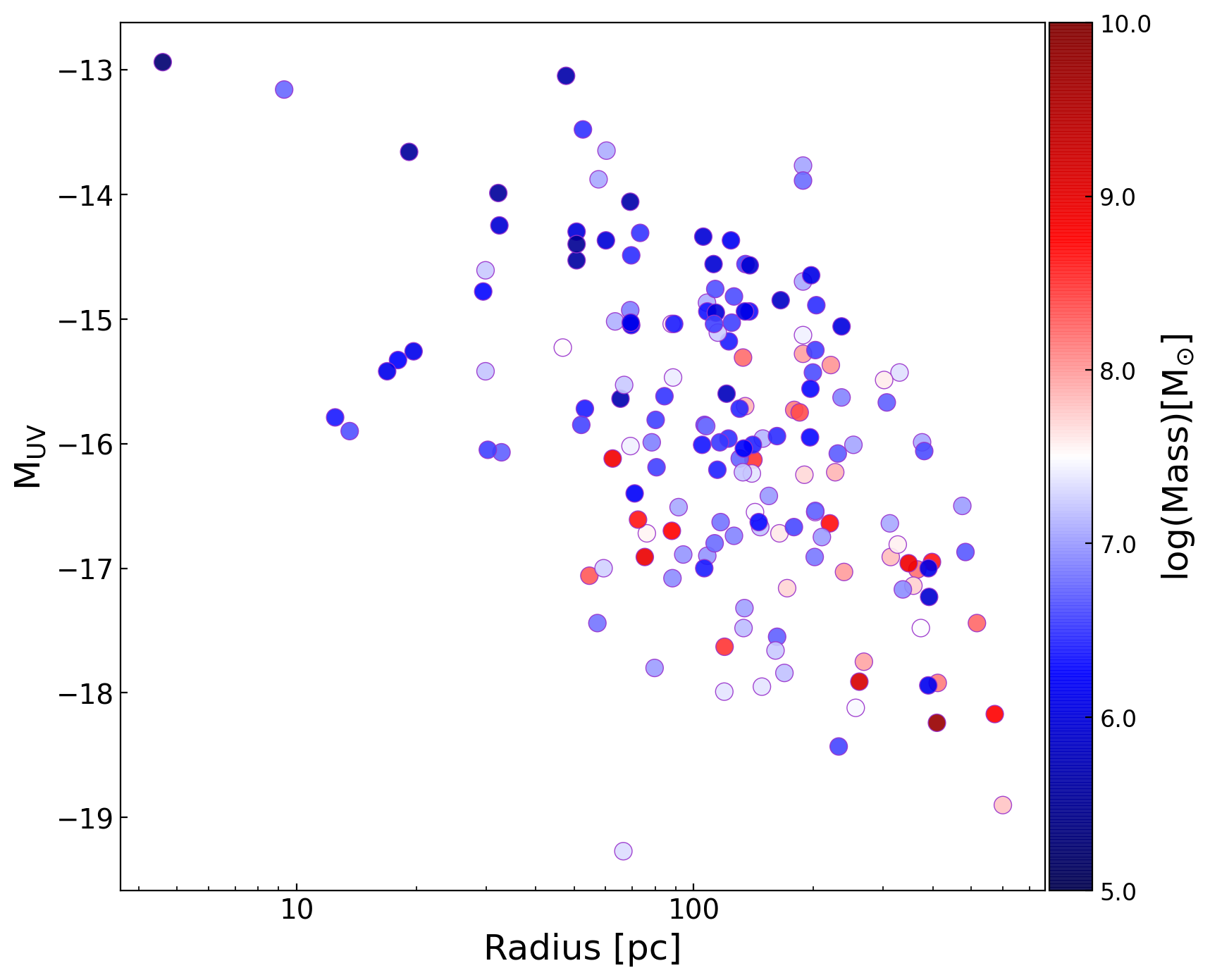}}
    \hspace{-0.01cm}\vspace{-0.05cm}

    \subfloat{\includegraphics[width=.4\textwidth]{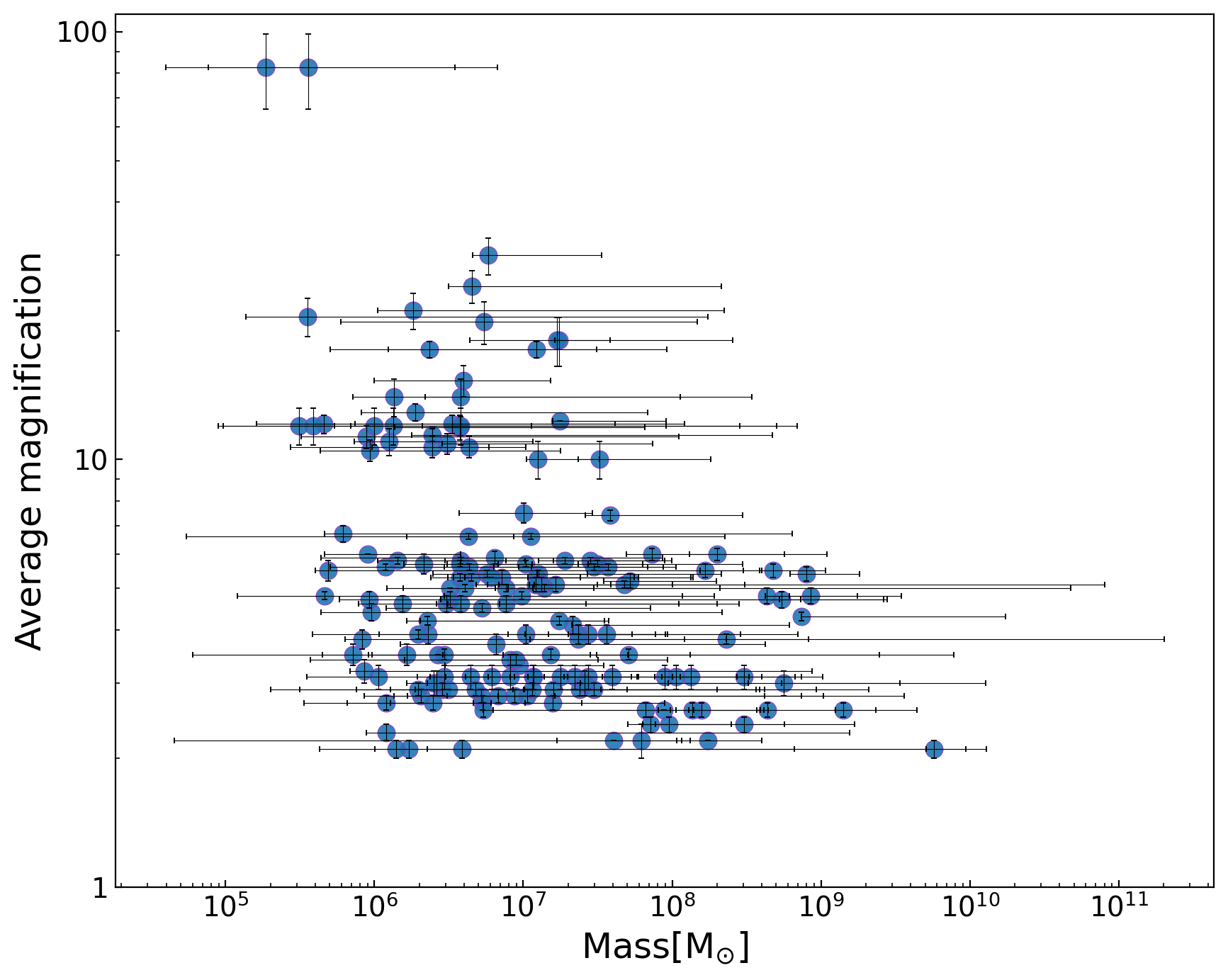}}
    \hspace{-0.01cm}\vspace{-0.05cm}
    \subfloat{\includegraphics[width=.4\textwidth]{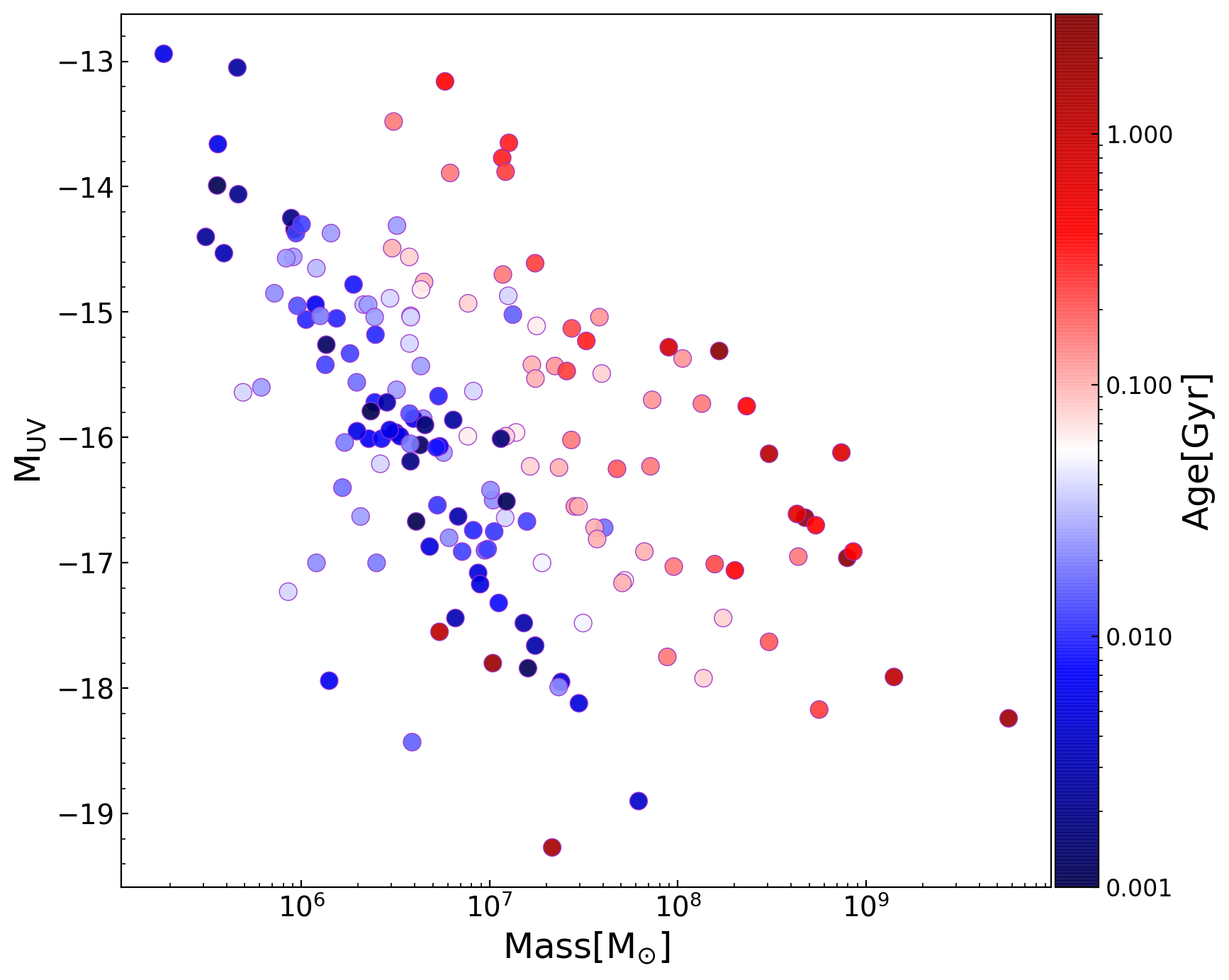}}
    \hspace{-0.01cm}\vspace{-0.05cm}

\caption{\textbf{Upper left}: Average magnification as a function of effective radius (R$_{\rm eff}$) in parsecs for the MACS~J0416 sample of star-forming clumps. For clumps for which we were unable to measure R$_{\rm eff}$, upper limits are provided (marked with orange-red diamonds). Clearly, highest magnifications correspond to clumps with smallest size. \textbf{Upper right}: Absolute magnitude as a function of effective radius color coded based on the stellar mass. A clear trend is noticeable, clumps with measured higher radius are brighter.  \textbf{Bottom left}: Average magnification as a function of stellar mass. No significant trend is observed as predicted from simulations. \textbf{Bottom right}: Absolute UV magnitude as a function of stellar mass where data is color coded based on age (Gyr) estimated from SED fitting. We can see that the detection threshold is more important than the resolution (magnification) when detecting less massive clumps.}
\label{fig:mass-size-mag-magnification}
\end{center}
\end{figure*}

Results from hydrodynamic simulations suggest that measured clump sizes depend on the spatial resolution of the observation \citep[e.g.][]{Tamburello2015,Tamburello2017, Faure2021}. 
This is consistent with our results: the sizes of clumps detected in MACS~J0416 decrease as lensing magnification increases and the number of detected clumps increase with magnification, Figure \ref{fig:20.0c}. 
In the most extreme cases where $\mu>50$ is reached, intrinsic sizes of $\sim$5 pc are measured, see Figure \ref{fig:mass-size-mag-magnification}. 
Clumps with sizes of tens of parsecs are also reported in the literature, for example \citet{Vanzella2021_sun, Calura2021} observed lensed clumps with sizes of few parsecs at $z=2.37$, and \cite{Vanzell2019MNRAS_clumps} measured sizes of a few tens of parsecs at $z\sim6$ (the magnification is $\mu>50$ in both cases).
In field instead (e.g $z\sim2$), where the spatial resolution of the observations is coarser, objects have sizes $\sim1$ kpc \citep{Schreiber2018}.

The median size of clumps in our sample is 126 pc, with a standard deviation $\sigma \sim$ 117 pc. 
This is comparable with the median size of clumps found in simulations \citep{Tamburello2017}, although our sample extends to even smaller sizes (e.g., R$_\mathrm{eff} < 100$ pc). 
Simulations with higher resolution than currently available are likely needed to compare with our highest magnification clumps.
On the other hand, in our sample $23\%$ of the clumps have effective radii $200 \lesssim \mathrm{R_{eff}} \lesssim 500$ pc. 
Such clumps have magnification $\mu \sim 2 - 11$ and are resolved. 
These large clumps are completely missing in the simulations by \cite{Tamburello2017}, possibly due to the relatively low initial molecular gas fraction of the host galaxy ($\sim\! 25$\%) and strong feedback recipe used in the simulations \citep{Fensch2021}. 
Large clumps are instead found in simulations adopting larger molecular gas fractions ($\sim\! 50$\%) and moderate feedback recipes \citep{Bournaud2014}. 
This indicates that to have a proper comparison between observations and simulations, spatial resolution, as well as the initial conditions (e.g., molecular gas fraction, feedback) for simulations are key.

In state-of-the art cosmological hydrodynamic simulations performed at very high resolution \citep[i.e. dark matter particle mass resolution of  $\sim 1000~\rm M_{\odot}$ and, in some cases, pc-scale softening length for baryons,][]{Ma2020}, bound and dense clumps in gas-rich high-redshift galaxies have been identified, with sizes and masses below 100 pc, consistent with the typical values of star clusters \citep{Kim2018, Phipps2020} and with the values measured in our sample. 
A systematic comparison between the properties of our clumps 
and the ones of current state-of-the art simulations is deferred to a future work. 

While the measured size of clumps clearly depends on the resolution of the observations, their stellar mass estimates seem to rather depend on the depth and detection threshold of the data \citep{Dessauges-Zavadsky2017}.
We show in Figure \ref{fig:mass-size-mag-magnification} the relation between the stellar masses of the clumps and their magnification. We only find a weak correlation (evaluated Pearson coefficient $\rm R=-0.11$). 
The trends that we are observing between the measured properties (mass, size) and the observational resolution and sensitivity are also consistent with results from simulations \citep{Tamburello2017}. 
In fact, \cite{Tamburello2017} report that when changing resolution from 100 pc to 1 kpc, at a fixed detection threshold, the estimated stellar masses are a factor $\sim2$ larger. Instead when increasing the sensitivity by a factor 10 the measured stellar masses change by one order of magnitude.

\section{Scaling relations} \label{scaling relations}

Various fundamental scaling relations have been used to unveil and understand relations between physical properties (e.g. between mass and size), of various dynamically evolved stellar systems \citep[i.e.][]{Misgeld&Hilker2011, Norris2014}.
However, the majority of such studies focus on stellar systems located in the nearby Universe, especially when faint and small sources are targeted. 
Thanks to lensing, we can extend scaling relations populating the low-mass, low-SFR, and small size portion of the parameter space with high redshift ($2\lesssim z \lesssim 6.2$) objects.

\subsection{Mass -- stellar surface mass density relation }
\label{subsec:mass_surf_density_rel}
\begin{figure*}
  \centering
  \includegraphics[width=15cm, height=8.78cm]{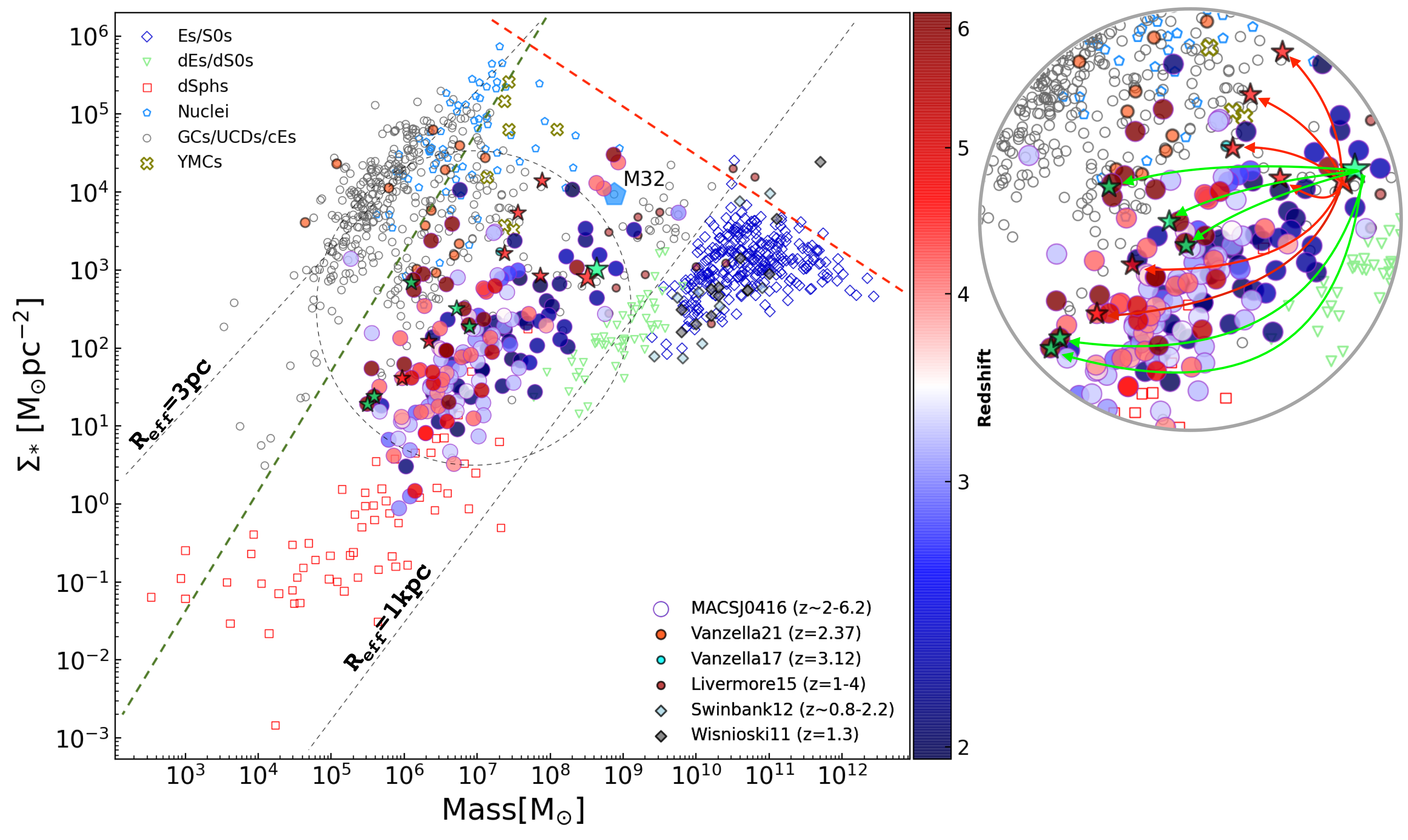}
  \caption{Mass - stellar surface density relation compared to different systems and samples from the literature. The lensed samples (from literature) are marked with circles while field samples are marked with diamonds. Our MACSJ0416 sample is represented by filled circles color coded with redshifts. The comparison sample of sources from the local Universe is represented with open symbols adopted from \citet{Norris2014}. 
  The stellar surface density is calculated following \citet{Misgeld&Hilker2011}. The red dashed line indicates the 'zone of avoidance' which define maximum stellar mass of the presented systems for a given R$_{\rm eff}$, the green dashed line corresponds to a relaxation time equal to the Hubble time and the black dashed lines trace systems with R$_{\rm eff}=1$ pc and R$_{\rm eff}=1$ kpc. The dashed circle marks the zoomed region of the figure shown to the right, where 
  we show the location of the two sources ID106b (big red star) and ID124c (big green star) measured as single objects in the {\tt ASTRODEEP} catalog. The red and green arrows show the positions of their internal clumps (see Fig~\ref{galfit}), when considered individually, also marked with small red and green stars (see text for more details).}
  \label{fig:mass-stellar_surface_density}
\end{figure*}

Our sample of clumps is particularly suited to examine the stellar-mass vs mass-surface density plane at high redshift  (Figure~\ref{fig:mass-stellar_surface_density}), where the variation among different stellar systems is even more evident than on the stellar-mass vs size plane shown in APPENDIX \ref{app:mass_size_rel}.
In this plane, two sequences have already been discussed in the literature.
One is populated by sources with large mass (M$_\star \gtrsim 10^9$ M$_\odot$) and $\Sigma_\star \sim 10 - 1000$ M$_\odot$ pc$^{-2}$, such as Es/S0s and dEs/dS0s galaxies. 
The other is instead populated by sources with smaller mass (M$_\star \lesssim 10^7$ M$_\odot$) and larger stellar mass surface density ($\Sigma_\star \sim 10^2 - 10^5$ M$_\odot$ pc$^{-2}$).
In between those two sequences, 
there is a region sparsely populated, with only a few nuclear star clusters (Nuclei) and the M32 galaxy nuclei. 
The bulk of the clumps of our sample is located in an intermediate region.
Simulations \citep[i.e][]{Pfeffer2013} and local observations \citep{Norris2014} suggest that stellar systems located in this portion of the parameter space may possibly undergo stripping processes, during which their outskirts are removed, leaving only a dense core.
In the process of the stripping, galaxies are expected to migrate from the sequence populated by dES/dS0s to the region occupied by compact and dense stellar systems.
This scenario could potentially explain the position of some clumps from our sample, particularly those systems where two or more clumps are detected.

Another possible explanation for the specific location of our clumps is that we may not be resolving clumps into individual stellar systems, despite the high magnification, so that they could rather be star-forming regions where small star clusters are blended due to insufficient resolution. 
With increasing resolution, we indeed expect them to be placed in the region occupied by local GCs, UCDs, and cEs.
An exemplification of this effect is  illustrated in Figure \ref{fig:mass-stellar_surface_density}
with two examples: system ID106b and ID124c, with delensed total stellar mass M$_\star = 3.3\times10^8\,\mathrm{M}_{\odot}$ and M$_\star = 3.6\times10^8\,\mathrm{M}_{\odot}$ (non delensed stellar masses are adopted from the {\tt ASTRODEEP} catalogue, where both sources are treated as a single object, \citealt{Merlin2016A&A}). 

Stellar masses and effective radii are measured as described in Section \ref{subsec:clump_properties}. 
For ID106b and ID124c,
we obtaine delensed values of R$_\mathrm{eff}\sim 253$ pc and 257 pc  respectively. 
When considering the full-resolution image (i.e., without smoothing), we detect 6 (5) clumps within ID106b (ID124c) owing to the high-magnification ($\mu\approx 11-12$, see Figure~\ref{galfit}). 
In this way, 
we notice from Figure \ref{fig:mass-stellar_surface_density} that the location of the individual clumps hosted by ID106b and ID124c is closer to the region occupied by local GCs and nuclei, when compared to the location of their hosting systems.
If we reasonably assume that such a resolution effect affects the majority of our clumps, 
this would explain why our sample occupies a sparsely populated region in between GCs/UCDs/cEs, YMCs, Nuclei, and Es/S0s, dEs/dS0s stellar systems.

\begin{figure*}
  \centering
  \includegraphics[width=15cm, height=10.4cm]{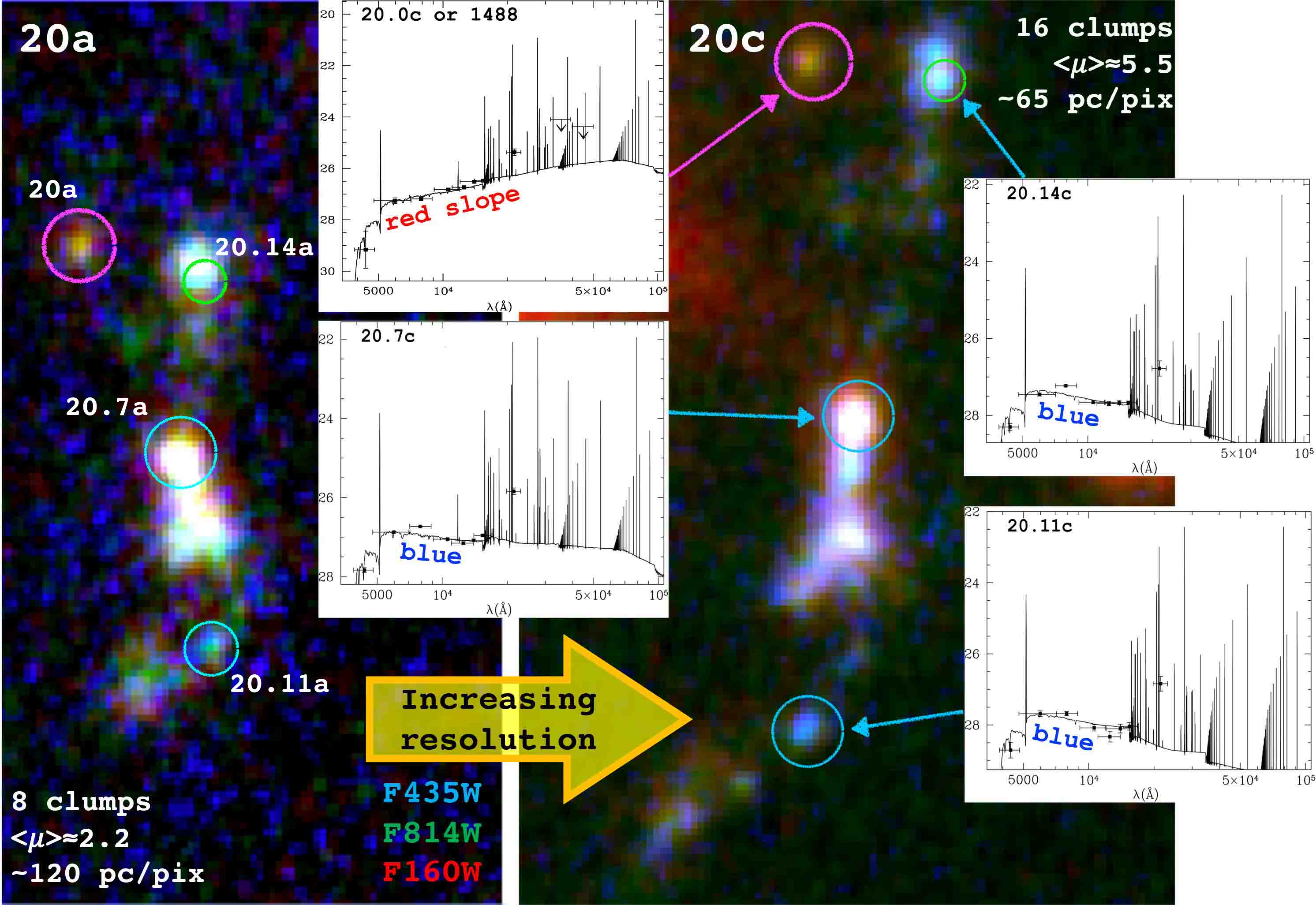}
  \caption{Two multiple images of system 20 ($z=3.222$) observed with different magnification in two multiple images of the same source: 20a (left, the least magnified) with $<\mu> \simeq 2.2$ and 20c with $<\mu> \simeq 5.5$ (right). The number of detected clumps increases  with the effective resolution (i.e., with increasing magnification). Four clumps are indicated with arrows with their best SED-fit solutions (insets). Three out of four appear blue star-forming regions with possible clear nebular emission dominating the K-band magnitude, whereas clump 20.0c (indicated with a magenta circle) shows a smoothly red shape suggesting it is relatively old, massive and gravitationally bound (see text for more details).}
  \label{fig:20.0c}
\end{figure*}

Interestingly, even though the clumps in our sample might have unresolved substructure, some of them could be gravitationally bound and considered as individual and stable star-forming regions that survive and potentially evolve over time.
An example is the clump 20.0c shown in Figure~\ref{fig:20.0c}. Such an object shows two multiple images with mean magnifications of 2 and 5.5. Both images show a nucleated morphology, with the most amplified being barely resolved and providing an estimated intrinsic effective radius of 100 pc. Differently from the other clumps populating the same system 20, clump 20.0c shows a smoothly red continuum with associated physical quantities compatible with being a relatively evolved and massive object, with an age of $\sim 40$ Myr and a de-lensed stellar mass of $\sim 10^{8}$ \msun. After including 68\% uncertainties, such values correspond to a relatively large stellar surface density $\Sigma_\star = 1500_{-800}^{+1000}$ M$_\odot$ pc$^{-2}$, with a dynamical age $\Pi = 2.5_{-1.9}^{+5.5}$ (with $\Pi$ defined as the ratio of the age and crossing time, see e.g., \citealt{Vanzella2021A&A}), suggesting 20.0c could be a gravitationally bound object.

\subsection{Mass -- SFR relation }
\label{subsec:mass_SFR_rel}

\begin{figure*}
\begin{center}
    \subfloat{\includegraphics[width=.49\textwidth]{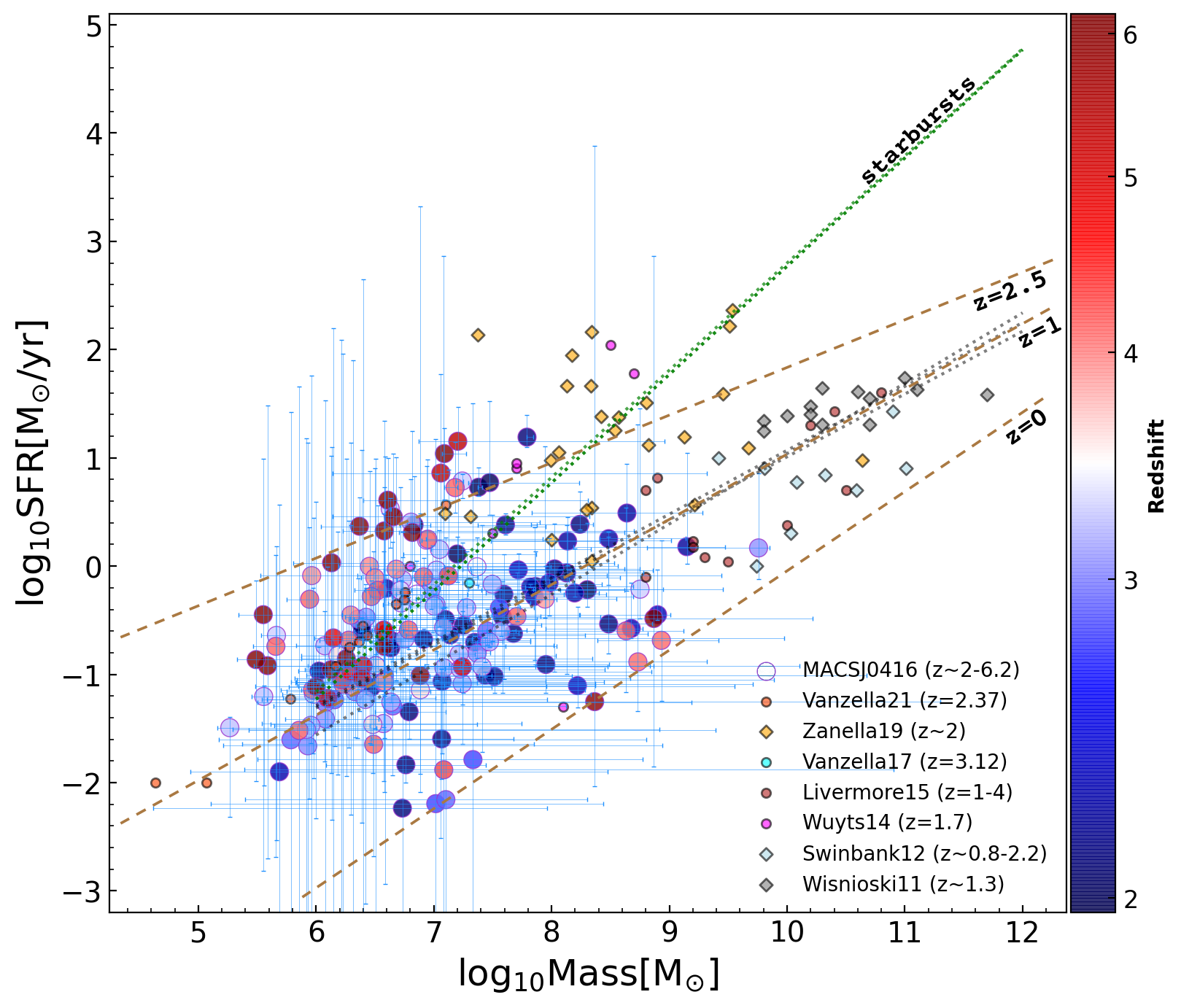}}
    \hspace{-0.01cm}\vspace{-0.05cm}
    \subfloat{\includegraphics[width=.49\textwidth]{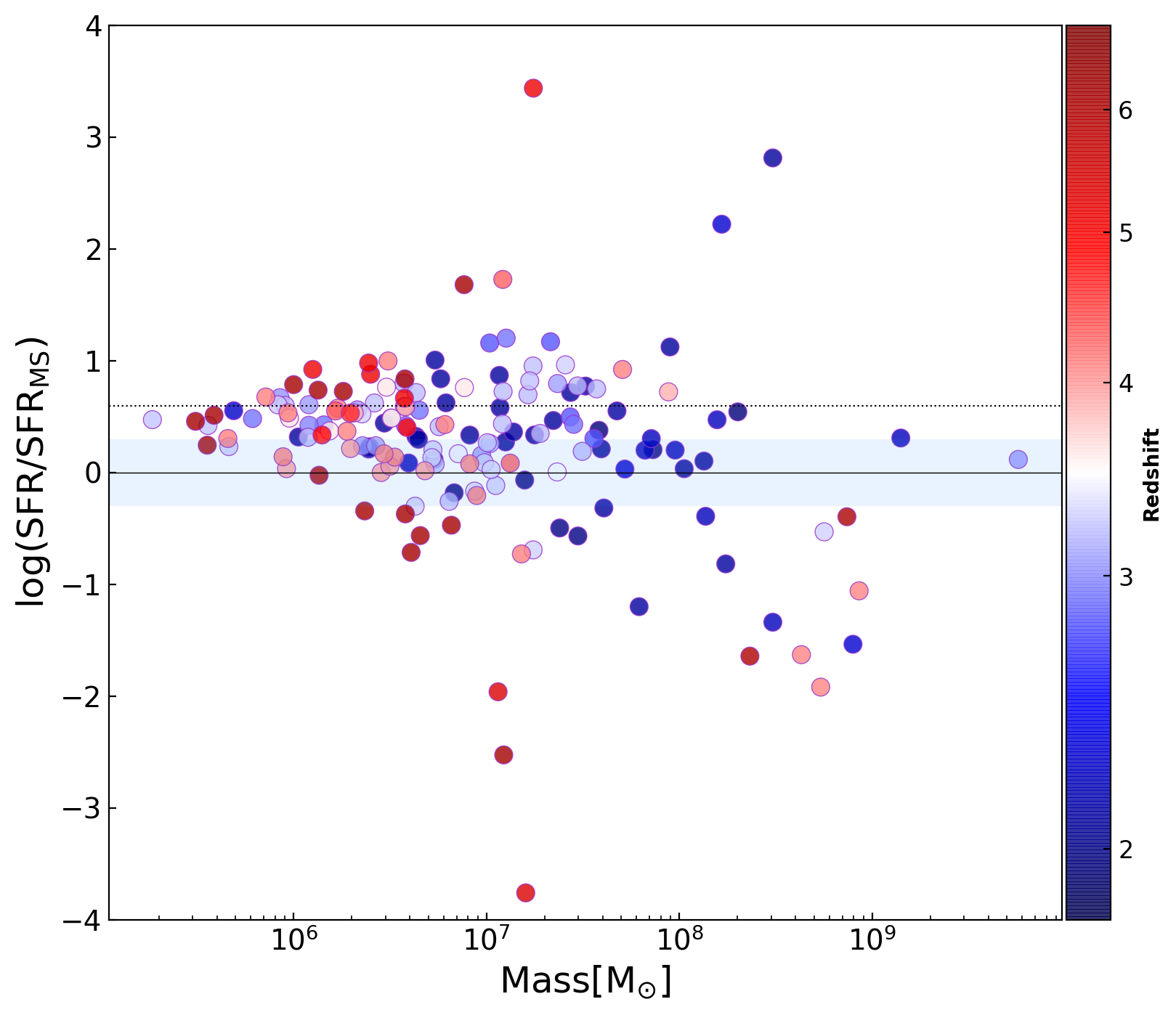}}
    \hspace{-0.01cm}\vspace{-0.05cm}
  
\caption{The mass--SFR relation. \textbf{Left panel}: The star formation rate as a function of stellar mass is shown for our sample of MACS~J0416 background clumps and compared with other lensed and field samples from the literature. The MACS~J0416 clumps are color coded based on their redshift (right y-axis). Lensed samples are indicated with circles while field samples are marked with diamonds. The dashed lines are taken from \citet{Whitaker2012} and represent main-sequence at $z=0,1$ and 2.5. The three dotted black lines (green lines Starburst) are adopted from \citet{Rinaldi2021} and represent main-sequence galaxies (Starbursts) in three redshift bins $z=2.8-4$, 4--5 and 5--6.5. Our study extends the mass--SFR relation to low mass stellar system down to $\sim10^5$\(\textup{M}_\odot\), entering the regime of the GCs, Nuclei, etc.
\textbf{Right panel}: The same relation is shown normalized to the SFR of the main sequence at varying redshifts derived by \citealt{Santini2017}. Only the MACS~J0416 sample is shown and color-coded by redshift. The blue shaded region and the dotted horizontal line are adopted from \citet{Zanella2019} marking the dispersion of the mass-SFR relation (0.3dex) and starbursts, respectively. Shown uncertainties for mass and SFR are 16 and 84 percentile range derived from SED fitting.}
\label{fig:mass-sfr}
\end{center}
\end{figure*}

We stress that unlike other studies of the main-sequence high-redshift galaxies lensed by MACS~J0416 (e.g., \citealt{Rinaldi2021}), we do not consider integrated galaxies, but rather make an effort to deblend clumps from their host galaxies to investigate the star formation mode of individual star-forming regions at high-$z$. 
A tight correlation between the SFR and M$_\star$ exists for local and high-redshift star-forming galaxies, the so-called "main-sequence" \citep[e.g][]{Brinchmann2004, Noeske2007, Rodighiero2010,Sparre2015, Santini2017}.
In Figure \ref{fig:mass-sfr}, we show the location of our sample of individual clumps in the SFR-M$_\star$ plane and compare it with the other published samples. 
Thanks to lensing magnification, our sample probes the low-SFR and low-M$_\star$ region, down to M$_\star \sim10^{5}\,\mathrm{M}_{\odot}$, where only a few other lensed sources from the literature are located (e.g., the Sunburst clumps at $z=2.37$, \citealt{Vanzella2021_sun}).

In Figure \ref{fig:mass-sfr}, we also show the normalized SFR ($\rm log(SFR/SFR_{MS})$) versus the stellar mass, obtained after dividing the SFR by the expected main-sequence SFR ($SFR_{MS}$)\footnote{The main sequence star-formation rate is adopted from \citep{Santini2017} and it has been derived only for the systems with stellar masses $\gtrsim 10^8\, {\rm M}_\odot$. } at the redshift of each clump. 
For this purpose, we adopt the main-sequence determination at varying redshifts by \citealt{Santini2017}, which was determined for stellar masses $\gtrsim 10^8\, {\rm M}_\odot$.
We note that $\sim55\%$ of the sample lies above the main-sequence, with 30\% of the clumps having SFR $\gtrsim 0.3$ dex higher than SFR$_\mathrm{MS}$, a region commonly populated by highly starbursting galaxies \citep{Rodighiero2010}. 
This is generally consistent with previous findings in the literature (e.g., \citealt{Guo2012}, \citealt{Wuyts2012}, \citealt{Zanella2015}, \citealt{Karman2017}, \citealt{Cibinel2017}, \citealt{Zanella2019}).

\subsection{Radius -- SFR relation}
\label{subsec:radius_SFR_rel}

\begin{figure*}
  \centering
  \includegraphics[width=12cm, height=8cm]{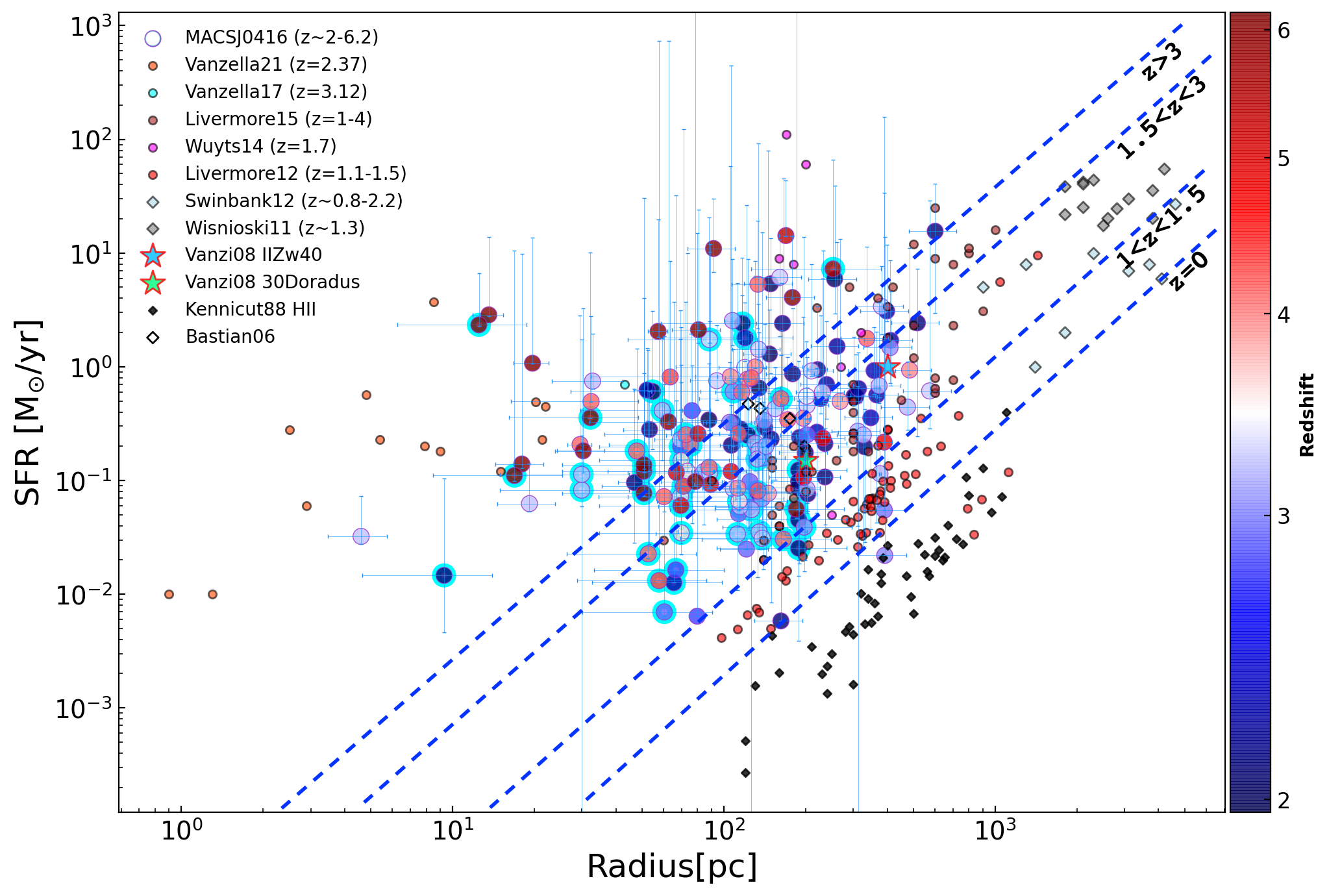}
  \caption{The size - SFR relation. The clumps from our work are plotted as large filled circles color coded by redshift. The comparison sample from the literature is divided in three categories, clumps observed in lensed fields (filled circles) clumps observed in field (filled diamonds) and well known local HII regions IIZw40 (blue star), 30 Doradus (green star, \citealt{Vanzi2008}) and the Antennae galaxy (open black diamonds, \citealt{Bastian2006}). The sample of local HII regions is adopted from \citet{Kennicutt1988} and only those HII regions for which radius was available are plotted (46 in total), the SFR is evaluated from the $L_{\rm H\alpha}$ using the relation $\rm SFR(\mathrm{M}_{\odot} yr^{-1})=7.9\times 10^{-42} L_{\rm H\alpha}(\rm erg s^{-1})$ (\citealt{Kennicutt1998}). Dashed lines show fitted radius-SFR relations in four redshift bins adopted from \citet{Livermore2015}. Clumps with estimated R$_{\rm eff}$ upper limits are marked with cyan circles around them. Radius uncertainties are propagated analytically taking into account uncertainties due to tangential magnification and measurements with Galfit. The SFR uncertainties are 16 and 84 percentile range from SED fitting.}
  \label{fig:radius-sfr}
\end{figure*}

The size-SFR relation has been already well defined in the local Universe investigating HII regions in nearby spiral and irregular galaxies by \cite{Kennicutt1988}. 
On the other hand, outliers are mainly hosted by interacting systems as shown in the case of the Antennae galaxy \citep{Bastian2006}, and have been thoroughly investigated on a larger samples with DYNAMO \citep[][]{DYNAMO} and LARS \citep[][]{Messa2019LARS}.
In Figure \ref{fig:radius-sfr}, we show the relation between the size and star-formation rate of clumps, including samples across different redshifts \citep{Swinbank2009, Jones2010, Swinbank2012, Livermore2012, Livermore2015,Johnson2017}. 
We consider samples where SFRs was estimated through SED fitting (as for our sample) or from H$\alpha$, using the prescription by \cite{Kennicutt1988}.
Our sample and the Sunburst clumps \citep{Vanzella2021_sun}, despite being at $z \sim 2 - 6$, have sizes comparable to local HII regions and GCs \citep{Jones2010, Swinbank2009, Livermore2012, Johnson2017}, but they have SFRs $\sim 300$ times higher. 
They also have SFRs $\sim100$ times higher than clumps from \cite{Wuyts2014} and \cite{Livermore2015}, which are part of a sample of lensed systems at $z > 2$.
In particular, we compare our sample with two well-known HII regions from the local Universe: 30 Doradus and II Zw40 \citep{Vanzi2008}.  
One possible explanation for the high SFR measured in our sample is enhanced interactions, or larger gas reservoirs at high redshift.
Similar cases have been observed in the local Universe, for example, the Antennae galaxy which is a local merging system hosting six star-forming complexes whose SFR ranges from 0.2--1.4 M$_\odot$ yr$^{-1}$, significantly higher than other local star-forming regions.
Three of those complexes show signatures of Wolf-Rayet stars, implying young ages of $\sim5\,\rm Myr$ \citep{Bastian2006}.
Similar properties are observed in the Sunburst 5.1 knot presented in detail in \cite{Vanzella2021_sun} and in some of our clumps shown in Figure D.1 of \cite{Vanzella2021A&A}.
Other scenarios, such as the fragmentation of gas-rich discs, are potentially responsible for the observed high SFR among high redshift clumpy structures \citep{Noguchi1999, Dekel2009_2}.
During this process, cold gas cools becoming unstable and the galactic disc starts to fragments and forms clumpy structures.
Further on, such newly formed structures lead to the increased SFR as discussed in i.e.
\cite{Immeli2004}.

\section{Summary and Conclusions} \label{Summary}

In this paper, we have presented a sample of 166 gravitationally lensed clumps spanning the spectroscopic redshift interval $2\lesssim z\lesssim 6.2$, magnified by a factor $\mu \sim 2 - 82$ and located behind the cosmic lens HFF~MACS~J0416 field.
This is the largest sample of lensed clumps currently available at these redshifts for which we have measured physical sizes and stellar masses, which critically benefits from spectroscopic observations of the MUSE Deep Lensed Field  \citep{Vanzella2021A&A}.
We carefully derived the effective radius for each clump and total magnitude in the reference filter (UV rest-frame) and inferred physical properties through SED fitting over seven \textit{HST} bands, including the ground-based VLT/HAWKI Ks. To this end, we take advantage of our high-precision cluster lens model, which is based on currently the largest number of spectroscopic multiple images, thus deriving intrinsic sizes, luminosities, SFR, and stellar masses. 
Owing to lensing magnification and magnification, we probe stellar masses of clumps down to $10^{5}$ \msun\ and physical scales as small as $R_{\rm eff}\sim 10$~pc.

We investigate the mass-SFR, mass-$\Sigma_{\star}$, and size-SFR scaling relations and compare them with several spectroscopic low and high-redshift samples from the literature (from lensed systems and field). 
The main results of our work are summarized as follows:
\begin{enumerate}

\item [i)] Gravitational lensing allows us to probe sizes in the range $10 \lesssim R_\mathrm{eff} \lesssim 500$ pc and stellar masses in the range $10^5 \lesssim \rm M_\star/\rm M_\odot \lesssim 10^9$.   
As expected, a linear $\mu$ - $R_\mathrm{eff}$ relation is observed, with the most magnified clumps being also the smallest. The properties ($\rm R_\mathrm{eff}$, M$_\star$) of our observed clumps are comparable to those of high-resolution simulations, 
although they cover spatial resolutions that such simulations do not achieve (e.g. $\rm R_\mathrm{eff}$ of tens of pc). 
Stellar ages estimated for our clumps spans from 1 Myr to 2.6 Gyr, 
suggesting that clumps could survive longer than estimated by current simulations.

\item [ii)] Statistically, at the given angular resolution provided by \textit{HST}, whenever magnification increases (the effective resolution increases), new individual star-forming clumps emerge. 
This is observed specifically on multiple images of the same system experiencing different magnifications. 
AO-assisted instrumentation achieving better spatial resolution is needed to understand whether such a trend continues down to single star clusters scale ($ \lesssim 30$ pc). 

\item [iii)] No clumps with radii R$_\mathrm{e} \gtrsim 500$ pc (or kpc size clumps) are observed in our sample, while such clumps are usually observed in field which cover larger survey volumes.

\item [iv)] In the stellar mass surface density - stellar mass plane our clumps occupy the region between galaxies and GCs ($\gtrsim10^{8} \rm{M}_\odot$), where only few sources have been found so far \citep[e.g. M32 nuclei,][]{Norris2014}. While this may indicate a population of objects that bridge galaxies with GCs, it could also be an effect of resolution, as higher magnified clumps tend to lie closer to the GCs region.

\item [v)] We stress that the study of \cite{Santini2017} probes masses M$_\star\gtrsim10^{8}\rm{M}_\odot$, while our sample have masses down to $10^{5}\rm{M}_\odot$. However, it appears that after comparing our results with those from \cite{Santini2017} we notice that clumps in our sample are active sites of star formation, with 30\% of them being starbursting (i.e. they are located on average $\sim 0.6$ dex above the main-sequence at their redshift).
Therefore, it appears that the SFR of the clumps are higher when compared with local star-forming regions of higher masses.

\item [vi)] Currently, available simulations are unable to reproduce clumps with small sizes and masses at high redshifts as observed in our sample.

The physical properties of the high-redshift clumps remain relatively unexplored, and their role in the formation and evolution of the galaxies is not clear.
Shortly, the JWST telescope will help us to better constrain and understand the physical parameters of the clumpy formation detected in the distant galaxies by extending the probed wavelength range in the optical rest-frame (redward of the K-band) with both imaging and spectroscopy, reaching with NIRCam an expected PSF FWHM smaller than $0.2''$ up to 5$\mu m$. 

\end{enumerate}

\section*{Acknowledgements}

We would like to thank anonymous referee for very useful comments which gave us a possibility to improve our work and address several issues that were initially overlooked.
This project is partially funded by PRIM-MIUR 2017WSCC32 ``Zooming into dark matter and proto-galaxies with massive lensing clusters''. We acknowledge funding from the INAF for ``interventi aggiuntivi a sostegno della ricerca di main-stream'' (1.05.01.86.31).
We acknowledges support from
PRIN INAF 1.05.01.85.01. 
GBC acknowledges the Max Planck Society for financial support
through the Max Planck Research Group for S. H. Suyu and the academic
support from the German Centre for Cosmological Lensing. 
We would like to thanks developers of the open-source packages for Python used in this work: Matplotlib \citep{Matplotlib2007}, Pandas \citep{PANDAS2010data}, NumPy \citep{NUMPY2020array}. 
Also we would like to acknowledge TOPCAT \citep{TOPCAT2005}.

\section*{Data Availability}

The data used in this study will be shared upon reasonable request to the corresponding author.



\bibliographystyle{mnras}
\bibliography{bib} 




\appendix

\section{Mass -- size relation}
\label{app:mass_size_rel}

\begin{figure*}
\begin{center}
    \centering
  \includegraphics[width=10cm, height=8cm]{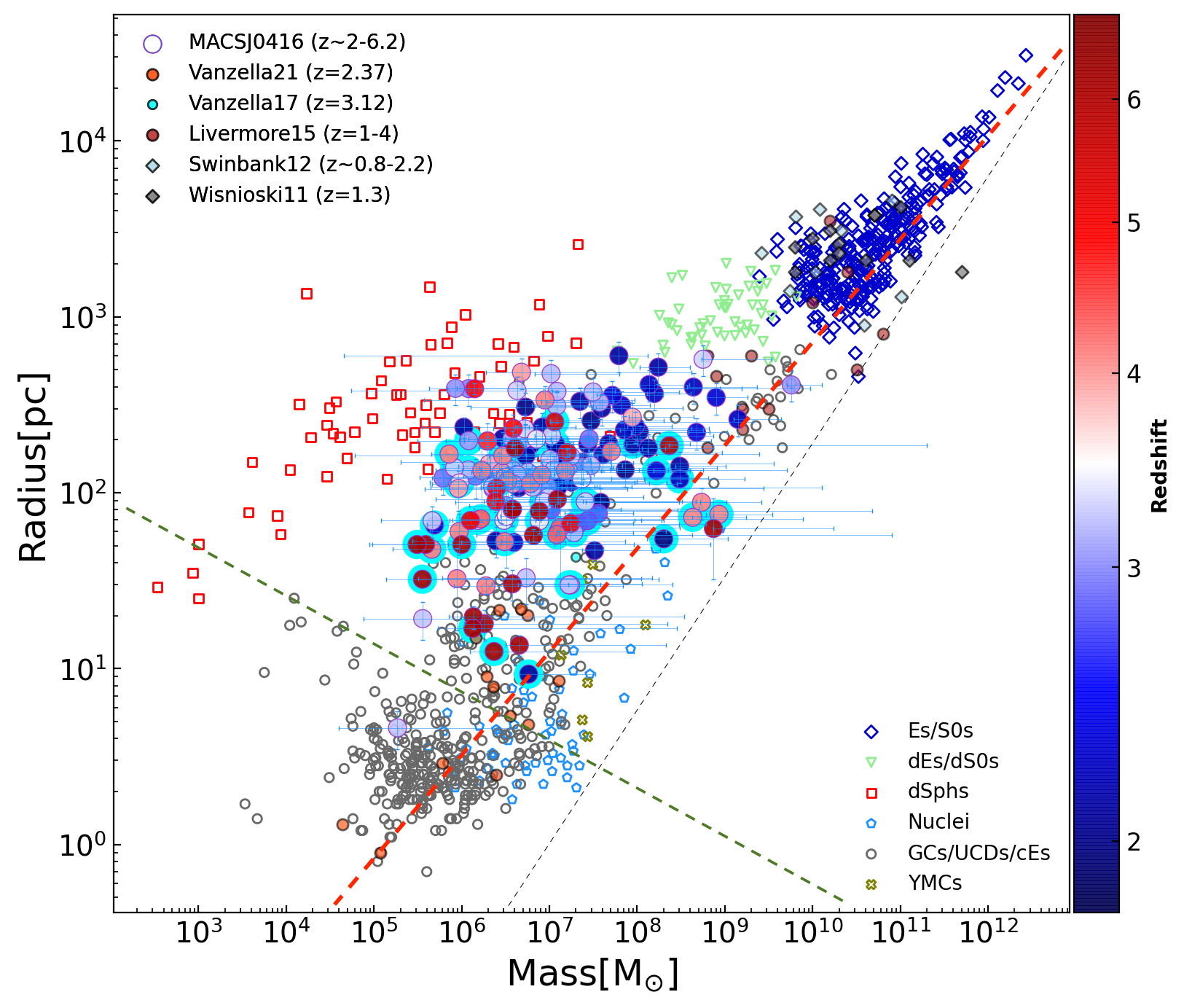}

\caption{The size--mass plane. Effective radius (R$_{\rm eff}$) in parsec is plotted as a function of stellar mass (M$_\star$) for a variety of stellar at high redshift and in the local Universe. The color coding is the same as in Figure \ref{fig:mass-stellar_surface_density}. Lensed samples include objects at higher redshift significantly smaller in mass and radius. The black dashed line represents 'zone of avoidance' corresponding to a maximum stellar mass for a particular radius (in local systems); the red dashed line is the size-mass relation from  \citet{Dabringhausen2008MNRAS}. The green dashed line, taken  from \citet{Misgeld&Hilker2011}, corresponds to a relaxation time equal to the Hubble time (separating collisional from collision-less systems).}
\label{fig:mass-size}
\end{center}
\end{figure*}

In Figure \ref{fig:mass-size}, we  present the mass-size relation which includes samples from the literature observed in field and lensed systems.
Recently, \cite{Bouwens2021} examined the same scaling relation focusing on low-luminosity objects with photometrically estimated redshifts at $6<z<8$ (333 sources in total) extracted from lensed fields.
Our MACS~J0416 sample is marked with big circles color-coded based on the redshift (right colorbar). 
Samples of clumps observed in lensed fields are marked with filled circles, while clumpy samples in field are marked with filled diamonds. 
The comparison sample of sources from the local Universe is represented by open symbols adopted from \cite{Norris2014}.
We note that MACS~J0416 sample probe stellar structures at $z>2$ below $10^{8}\,\rm{M}_\odot $ and 100 pc, thus bridging the gap between the low mass-size objects in the local Universe and their counterparts at high redshifts.
Recent studies have probed similar parameter space, (\cite{Vanzella2017(D11)}, \cite{Vanzella2021_sun}, \cite{Vanzell2019MNRAS_clumps}) by focusing on individual systems, whereas we  present here a statistically representative sample of compact stellar structures, which complements
studies in field and cover the upper right corner of the plot.

By comparing our spectroscopic sample with the one presented by \cite{Bouwens2021}, who  used different public lens models to estimate intrinsic properties, we conclude that both samples populate the region of star cluster complexes defined in \cite{Bouwens2021}.

Two different groups of stellar systems can be distinguished from Figure \ref{fig:mass-size}, galaxy-like systems (Es/S0s, dEs/dS0s, dSphs) and star clusters YMCs, GCs, UCDs. 
The MACS~J0416 lensed sample spans the $10^{5} - 10^{9}\,{\rm M}_\odot $ and $\sim$10--600 pc ranges, with the majority located above (or left) from the mass-size relation line \citep[red dashed line,][]{Dabringhausen2008MNRAS} filling a scarcely populated space in between dSphs and GCs, Nuclei, and YMCs.
In addition, it can be noted that clumps with lower average magnification are located closer to the dSphs, while those with higher magnification are closer to GCs, Nuclei, and YMCs stellar systems, suggesting that some of the clumps are not resolved sufficiently enough, so they may appear larger and more massive than in reality.
We also note that the most magnified systems in our MACS~J0416 sample populate the region in which the majority of GCs and Nuclei reside, indicating that we are starting to probe globular cluster like objects (or globular clusters precursors) at high redshift.

The black dashed line in Figure \ref{fig:mass-size} is the 'zone of avoidance' corresponding to a limiting stellar mass for a given effective radius  \citep{Misgeld&Hilker2011}. Interestingly, none of the clumps in our sample violate this limit, indicating that SF systems at high redshift follow the same relation.
Furthermore, only a few clumps are found below the green dashed line, which corresponds to a relaxation time equal to the Hubble time.
All objects below the green dashed line, a region populated by GCs and Nuclei, are collisional systems,  while collisionless systems lie above the green line, which include various types of galaxies and compact stellar systems.

Lastly, as suggested in \cite{Misgeld&Hilker2011}, it appears from Figure \ref{fig:mass-size} that the region in the mass interval $10^{6}\lesssim M_\star \lesssim 10^{7}\, \mathrm{M}_{\odot}$ divides stellar systems into two groups.
Objects with low mass, $<10^{6}\,\mathrm{M}_{\odot}$ (likely affected by dynamical evolution), do not seem to follow the mass-size relation , whereas massive systems do. 
However, low-mass systems are also  the most magnified clumps in our sample suggesting that any observed trend my be significantly affected by observation sensitivity and magnification, as discussed in Section \ref{subsec:sims}.

\section{Properties of the comparison sample from the literature}
\label{app:literature sample}

As shown in the main text, to better understand the physical properties of our sample and to put it in a broader context, we compare our results with other literature studies targeting clumps in the local Universe and at high redshift. 
In this work we focus only on spectroscopically-confirmed samples reported in the literature.
They can be divided into three categories: clumps observed in field at $z \gtrsim 1$, clumps observed in lensed fields at $z \gtrsim 1$, and evolved stellar systems in the local Universe. 
In the following, we will briefly describe the other samples observed in field at $z>1$ considered here.

Comparison spectroscopic samples observed in field at $z \gtrsim 1$:
\begin{itemize}
\item the sample from \cite{Wisnioski2011} includes 13 star-forming galaxies at $z \sim 1.3$ from the WiggleZ Dark Energy Survey, which was selected via ultraviolet and [OII] emission. Clumps have sizes R$_\mathrm{e} \sim 1-3$ kpc, stellar masses 
$\rm{M}_\star \sim 6.3\times 10^{9} - 5\times 10^{11}\,\rm{ M}_\odot$, and SFR $\sim 12 - 120\, \rm{M}_\odot yr^{-1}$. 
Stellar mass and SFR are estimated through SED fitting;

\item the sample from \cite{Swinbank2012} includes 9 H$\alpha$-selected galaxies at $z \sim 0.84 - 2.23$. Clumps have sizes R$_\mathrm{e} \sim$ 0.9 - 4.6kpc, stellar masses in the range M$_\star \sim 2.6\times10^{9} - 1\times10^{11}$\(\textup{M}_\odot\) and H$\alpha$-derived SFR $\sim 1-27\,\textup{M}_\odot yr^{-1}$;

\item the sample from \cite{Zanella2019} includes 53 star-forming galaxies at $z\sim 1-3$. We only consider the 25 clumps in this sample with 3$\sigma$ SFR estimates. They are unresolved (R$_\mathrm{e} < 500$ pc), have stellar masses in the range M$_\star \sim 10^{7}-10^{10}$\(\textup{M}_\odot\) and SFR $\sim 3-25\textup\,\textup{M}_\odot \textup yr^{-1}$.  Stellar mass were estimated through SED fitting, whereas the SFR considered here comes from dust-corrected rest-frame UV continuum estimates;

\end{itemize}

We also compare our sample with the the following spectroscopic samples drawn from lensed fields: 

\begin{itemize}

\item the sample from \cite{Livermore2012} which includes 8 gravitationally-lensed galaxies at $z=1-1.5$, where 57 clumps are detected. They have sizes in the range R$_\mathrm{e} \sim 98 - 1430$ pc and H$_\alpha$--based SFR $\sim 0.004-9.6\,\textup{M}_\odot \textup yr^{-1}$;

\item the sample from \cite{Wuyts2014} which includes one lensed young starburst galaxy at $z=1.7$ hosting 7 individual clumps with sizes R$_\mathrm{e} \sim 160 - 320$ pc and SFR $\sim 0.05-110\,\rm{M}_\odot yr^{-1}$; 

\item the sample from \cite{Livermore2015} includes 17 gravitationally-lensed galaxies at $z\sim1-4$. 
The size of the clumps is R$_\mathrm{e} \sim 0.17 - 3.5$ kpc and their SFR is derived from H$\alpha$ or H$\beta$ lines and ranges from 0.8 to $40\,\rm{M}_\odot yr^{-1}$. Their stellar masses are M$_\star \sim4\times10^{8}-6\times10^{10}\,\rm{M}_\odot$;

\item from the sample presented in \cite{Vanzella2017(D11)} we include one compact and extremely young ($<10$ Myr) clump at $z=3.1169$ with measured $\rm R_{e}\sim43$ pc, SFR$\sim0.7\,\rm{M}_\odot yr^{-1}$ and stellar mass of $20\times10^{6}\,\rm{M}_\odot$;

\item the sample from \cite{Vanzella2021_sun} includes one strongly lensed galaxy at $z\sim2.37$ (dubbed "Sunburst") where 13 star-forming clumps are detected. They have sizes R$_\mathrm{e} \sim 1 - 22$ pc, stellar masses in the range M$_\star \sim0.1\times10^{6} - 12.7\times10^{6}\,\rm{M}_\odot$ and SFR from $0.1 - 3.7\,\rm{M}_\odot yr^{-1}$;

\end{itemize}

Finally, the comparison samples of evolved stellar systems in the local Universe include: 
\begin{itemize}

\item elliptical and lenticular galaxies (Es/S0s), dwarf elliptical and lenticular galaxies (dEs/dS0s), dwarf spheroidals (dSphs), nuclear star clusters (Nuclei), globular clusters (GCs), ultra compact dwarfs (UCDs), compact ellipticals (cEs), young massive clusters (YMCs) and M32 nuclei. 
This entire sample is from \cite{Norris2014};

\item 30Doradus and IIZw40 from \citealt{Vanzi2008}; 

\item HII regions from \cite{Kennicutt1988}; 

\item star clusters from the Antennae galaxy \citep{Bastian2006}.

\end{itemize}

\section{The MACS~J0416 clumps}

\begin{figure*}
\begin{center}
    \centering
  \includegraphics[scale=0.33]{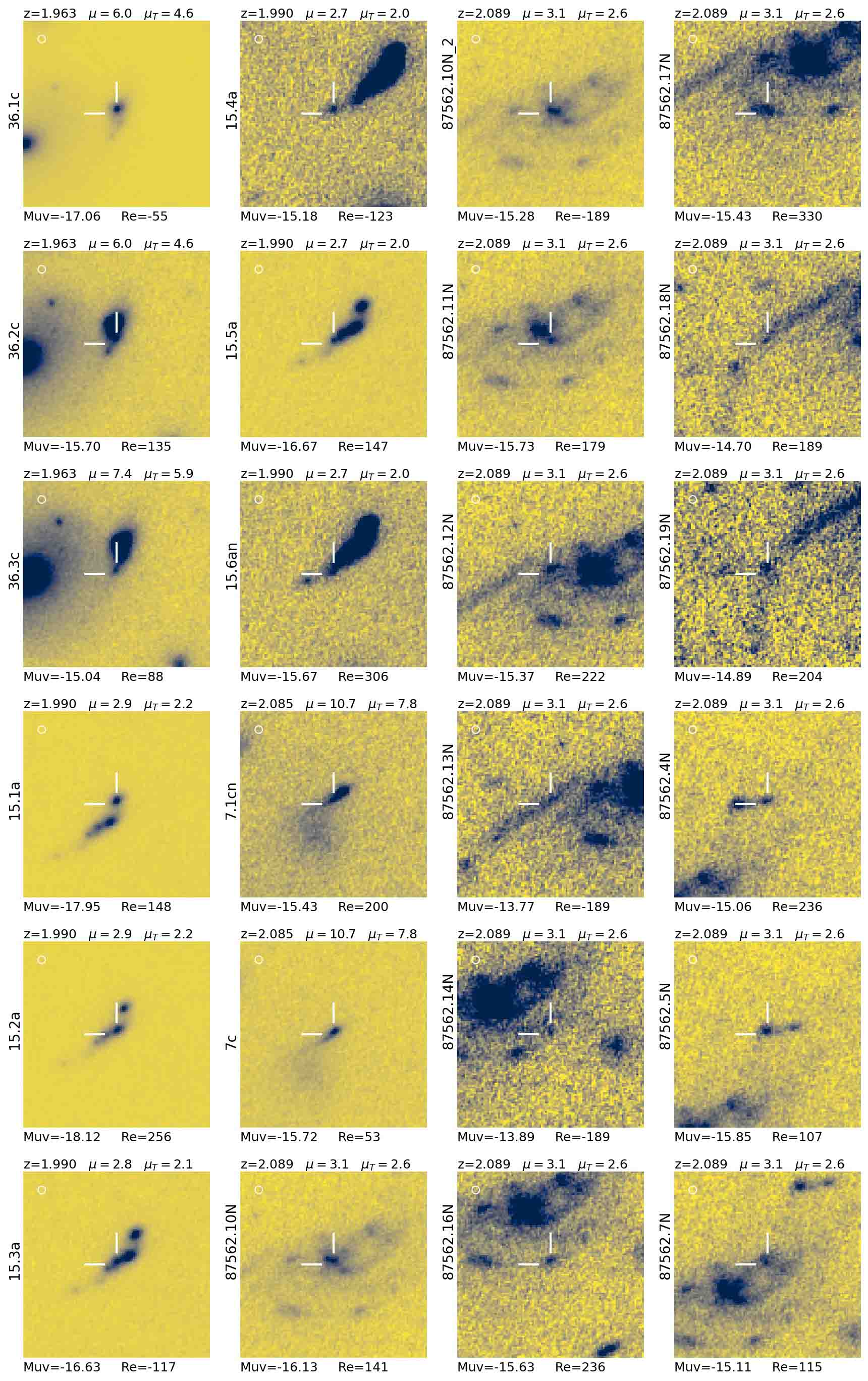}
\caption{The MACS~J0416 clumps from this work. We are showing an $HST$ F814W (F105W) $3''$ cutout images for detected clumps at $z<5$ ($z>5$). On the left y-axis is the clump ID, the upper x-axis shows redshift ($z$), average ($\mu$) and tangential ($\mu_{T}$) magnification, and the bottom x-axis shows absolute magnitude ($M_{\rm UV}$) and effective radius in parsecs (Re), where negative values for Re represent an upper limit. The circle in the upper left corner reports the F814W(F105W) PSF in case of $z < 5(> 5)$. The north on the cutout image is up and east is left. }
\label{fig:a2}
\end{center}
\end{figure*}

\begin{figure*}
\begin{center}
    \centering
  \includegraphics[scale=0.33]{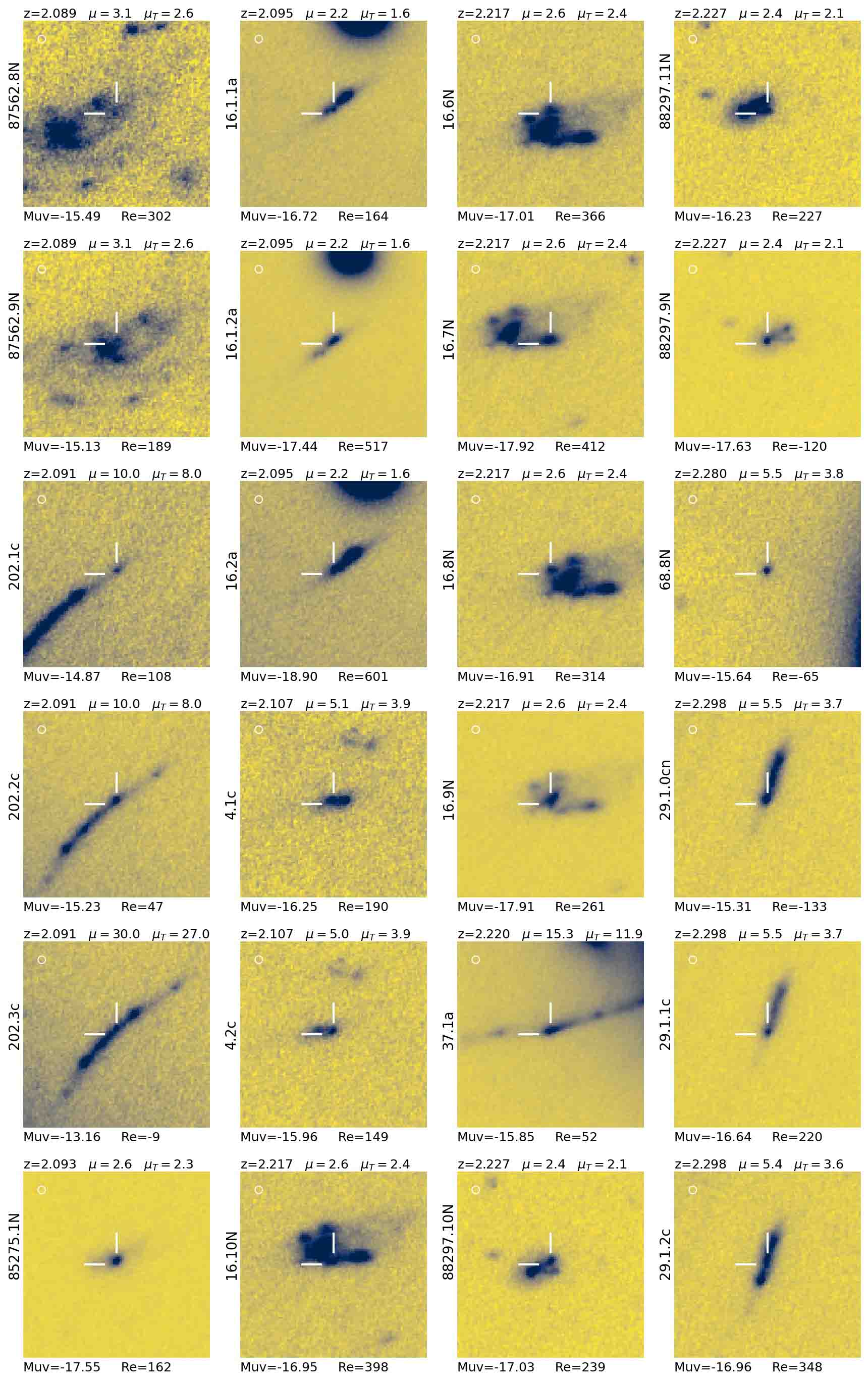}
\caption{Same as the \ref{fig:a2}.}
\label{a2.1}
\end{center}
\end{figure*}

\begin{figure*}
\begin{center}
    \centering
  \includegraphics[scale=0.33]{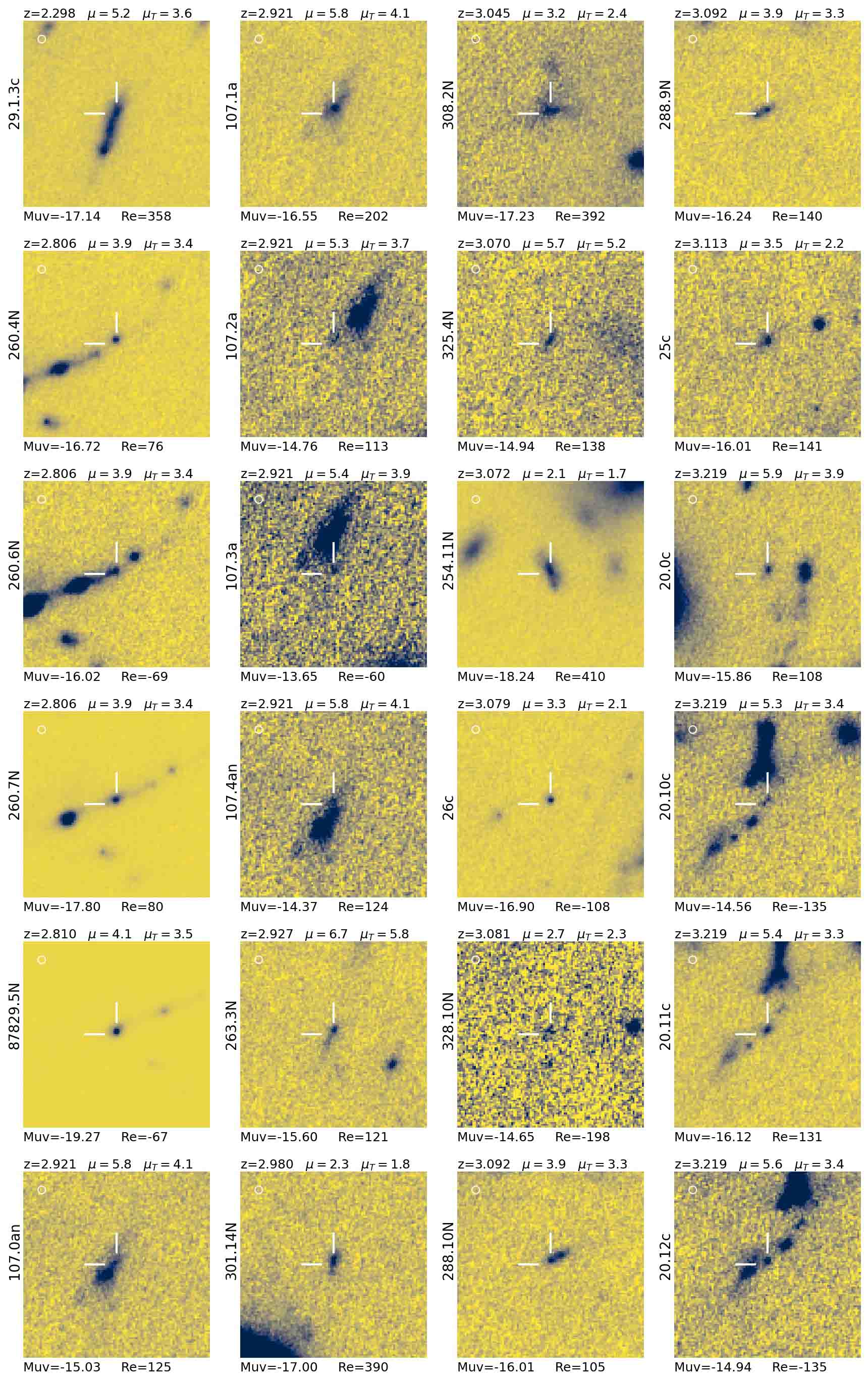}
\caption{Same as the \ref{fig:a2}.}
\label{a2.2}
\end{center}
\end{figure*}

\begin{figure*}
\begin{center}
    \centering
  \includegraphics[scale=0.33]{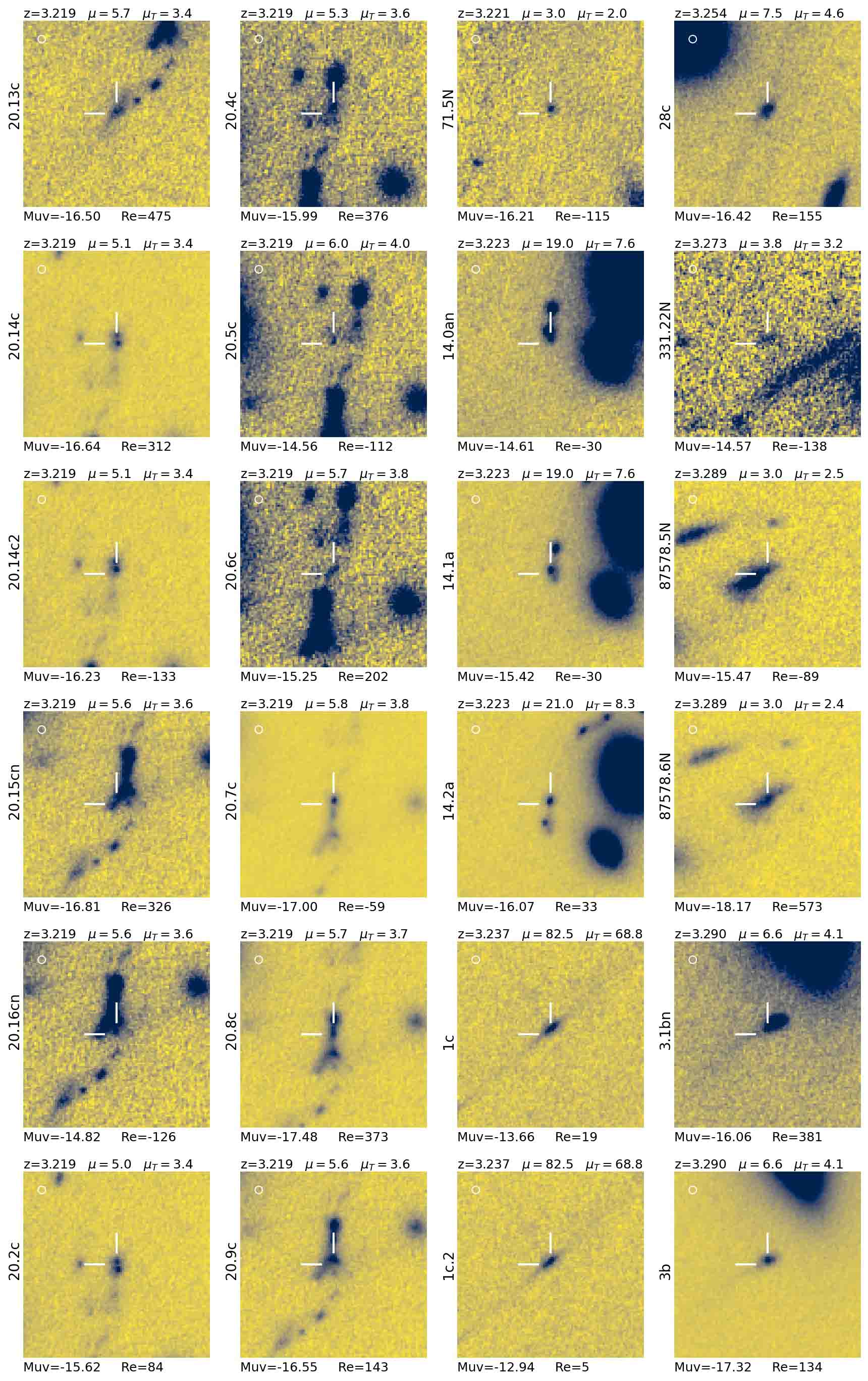}
\caption{Same as the \ref{fig:a2}.}
\label{a2.3}
\end{center}
\end{figure*}

\begin{figure*}
\begin{center}
    \centering
  \includegraphics[scale=0.33]{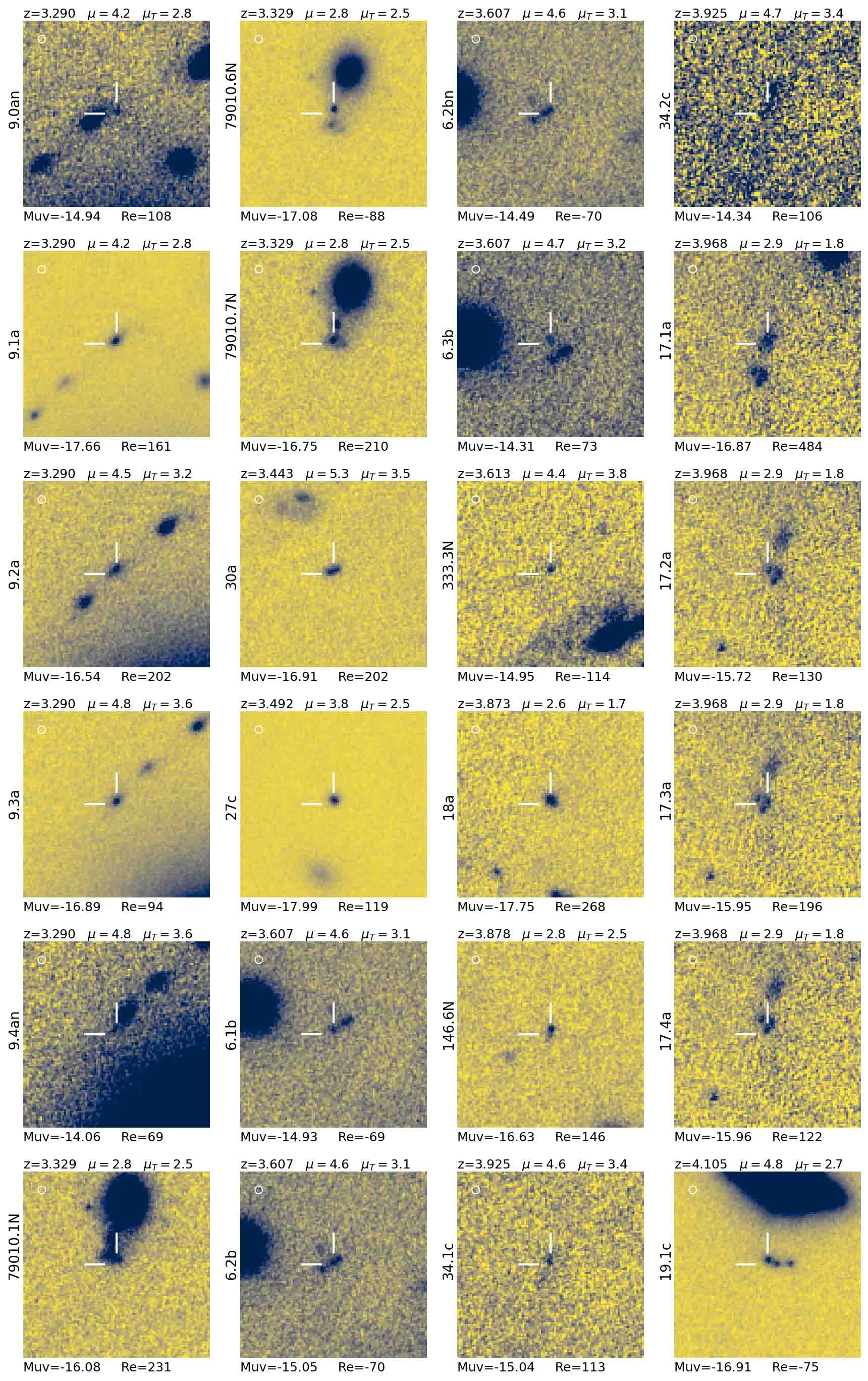}
\caption{Same as the \ref{fig:a2}.}
\label{a2.4}
\end{center}
\end{figure*}

\begin{figure*}
\begin{center}
    \centering
  \includegraphics[scale=0.33]{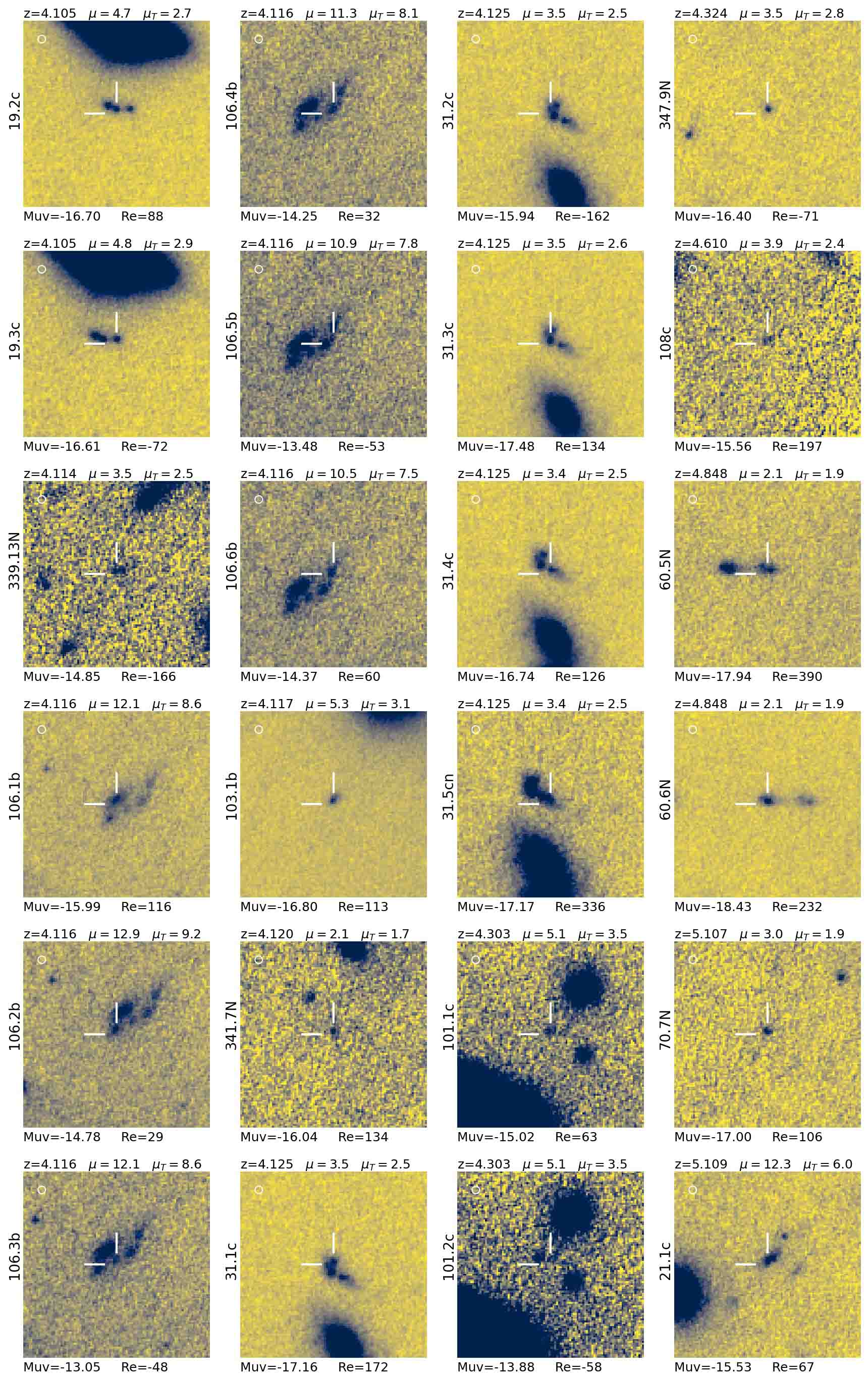}
\caption{Same as the \ref{fig:a2}.}
\label{a2.5}
\end{center}
\end{figure*}

\begin{figure*}
\begin{center}
    \centering
  \includegraphics[scale=0.33]{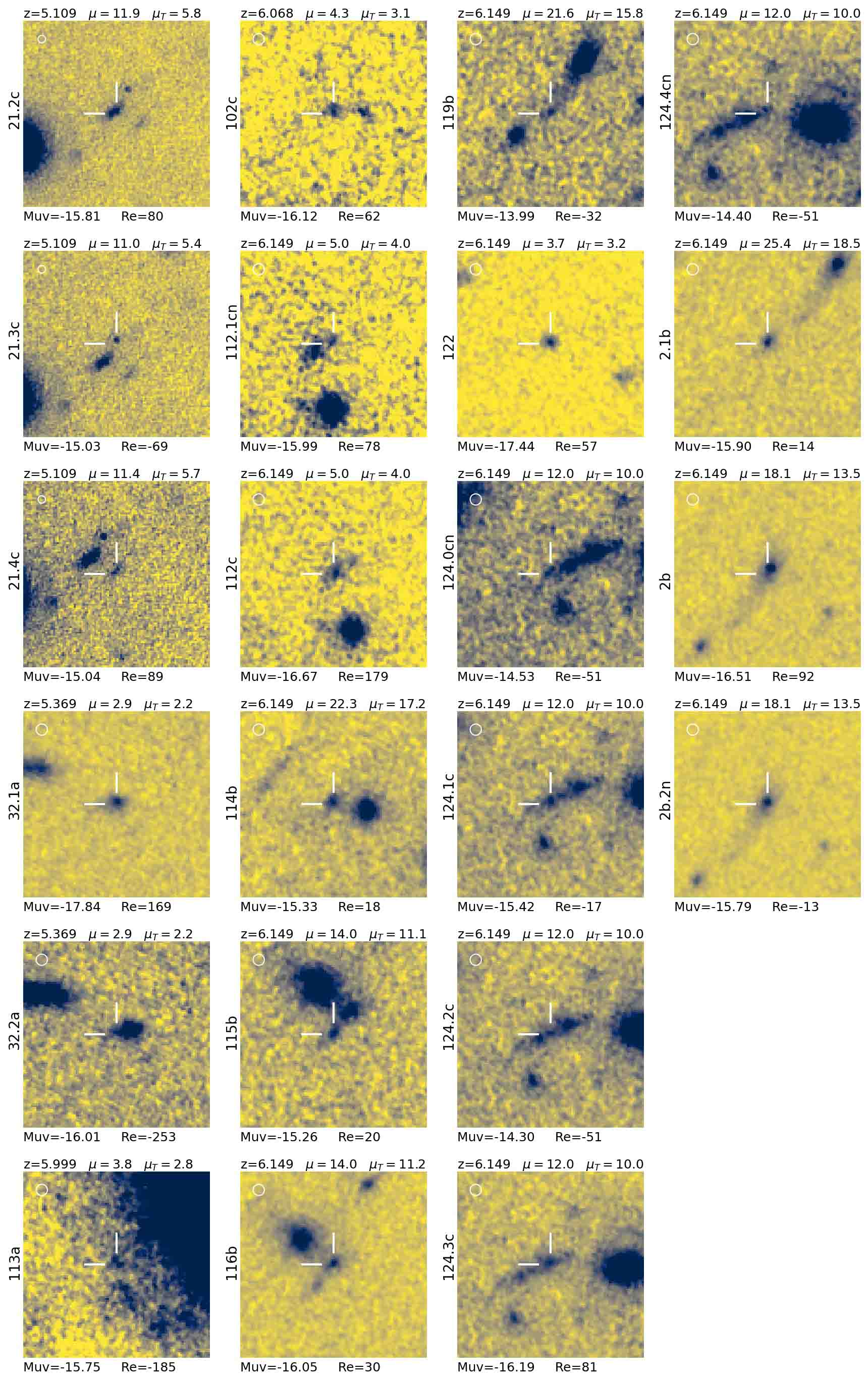}
\caption{Same as the \ref{fig:a2}.}
\label{a2.6}
\end{center}
\end{figure*}

Figures \ref{fig:a2}, \ref{a2.1}, \ref{a2.2}, \ref{a2.3}, \ref{a2.4}, \ref{a2.5}, \ref{a2.6} summarize the entire sample presented in this work by showing individual clumps as $3'' \times 3''$ cutout image (see captions of \ref{fig:a2} for more details).

\section{The robustness of the size measurement from {\tt GALFIT}}
\begin{figure*}
\begin{center}

    \subfloat{\includegraphics[ width=10cm, height=5cm]{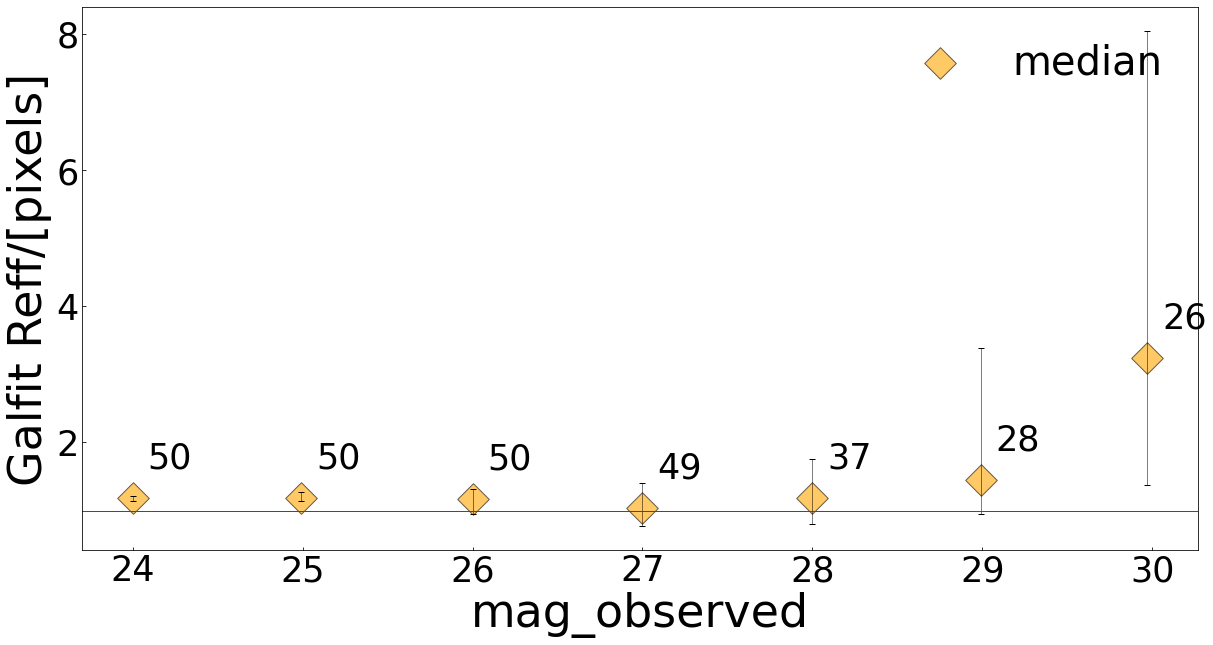}}
    \vspace{0.1cm}
    \subfloat{\includegraphics[ width=10cm, height=5cm]{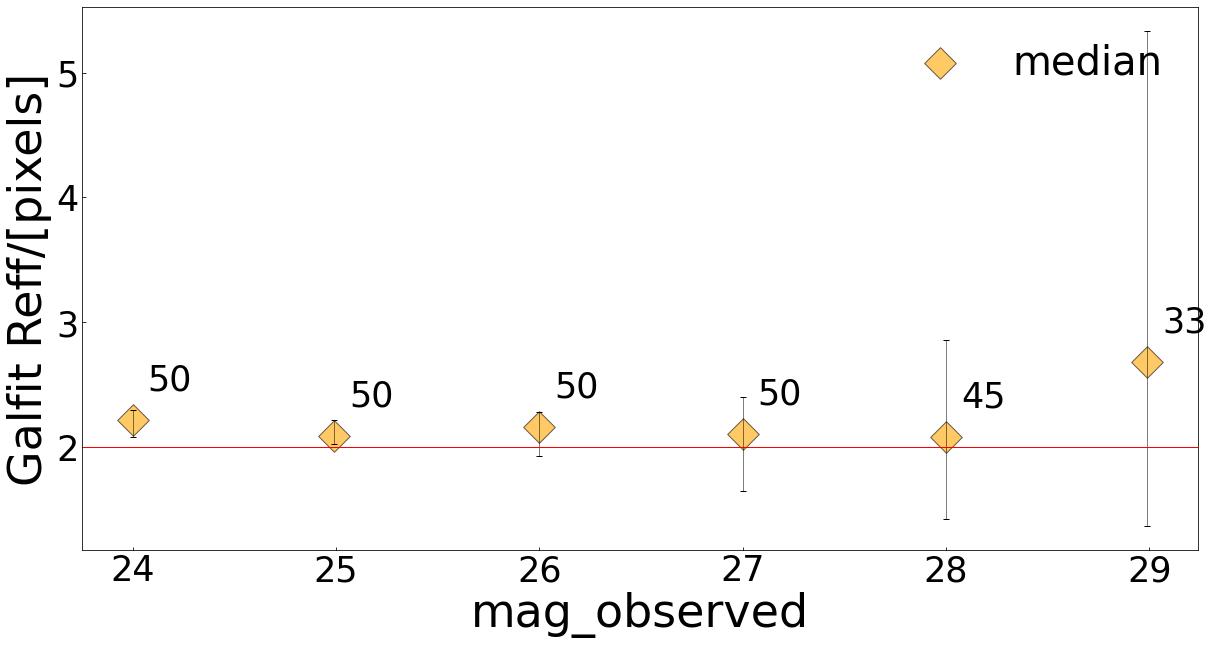}}
    \vspace{0.1cm}
    \subfloat{\includegraphics[ width=10cm, height=5cm]{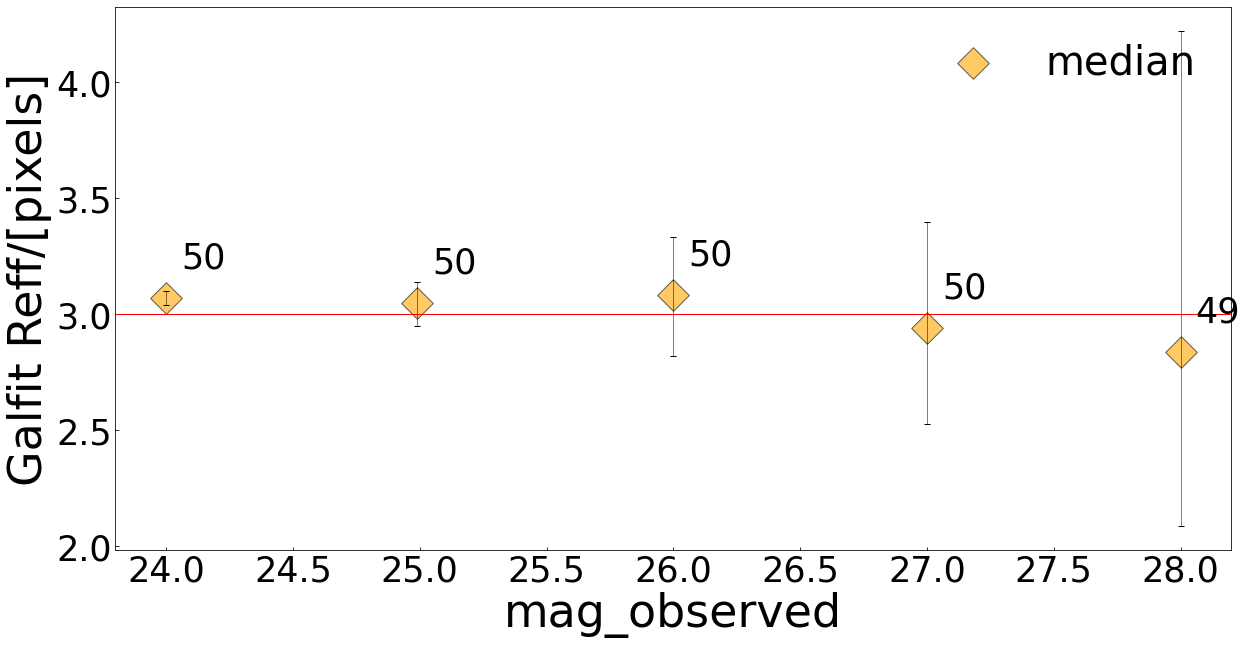}}
    \vspace{0.1cm}
    \subfloat{\includegraphics[ width=10cm, height=5cm]{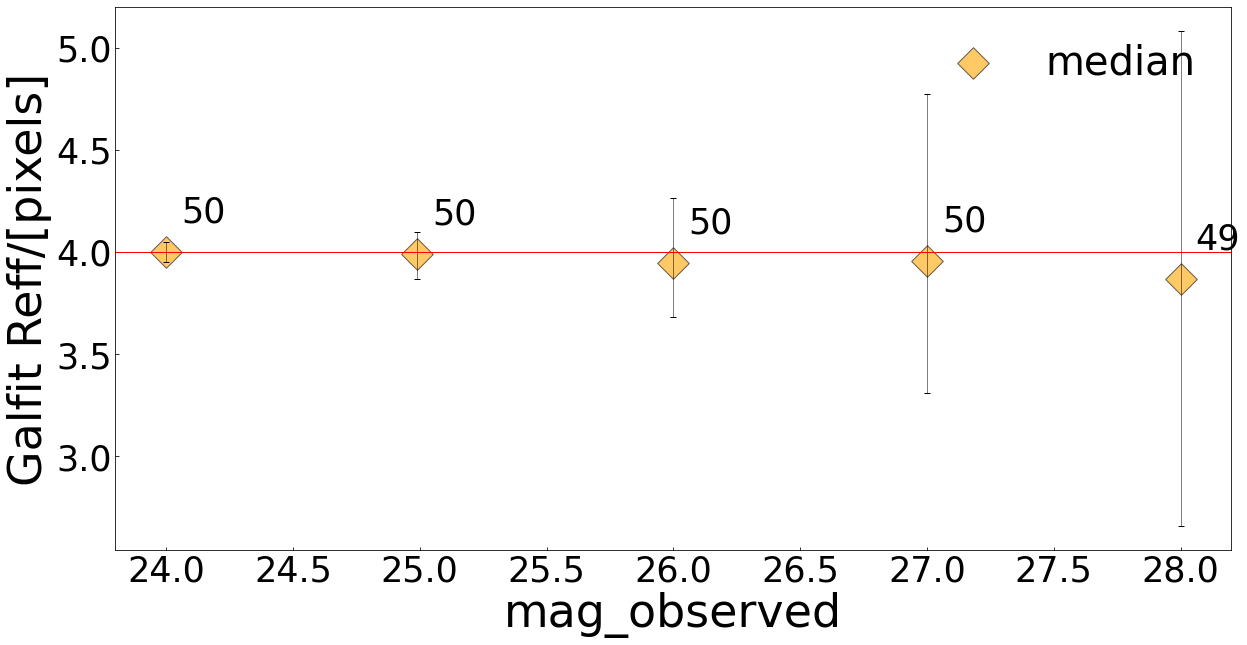}}
    
\caption{Results from {\tt GALFIT} runs on the simulated galaxies as described in the main text. 
On x-axis the range of magnitudes is reported and the y-axis shows the recovered R$_{\rm eff}$  in pixels (diamonds). Red horizontal line represents the input R$_{\rm eff}$ value in pixels with the numbers indicated near each diamond showing how many runs produced a {\tt GALFIT} solution (to be compared with the 50 MC).
Each diamond point represents the median R$_{\rm eff}$ recovered in the MC, along with the 16th and 84th percentiles (black error-bars). }
\label{D1}
\end{center}
\end{figure*}

\begin{figure*}
\begin{center}

    \subfloat{\includegraphics[ width=10cm, height=5cm]{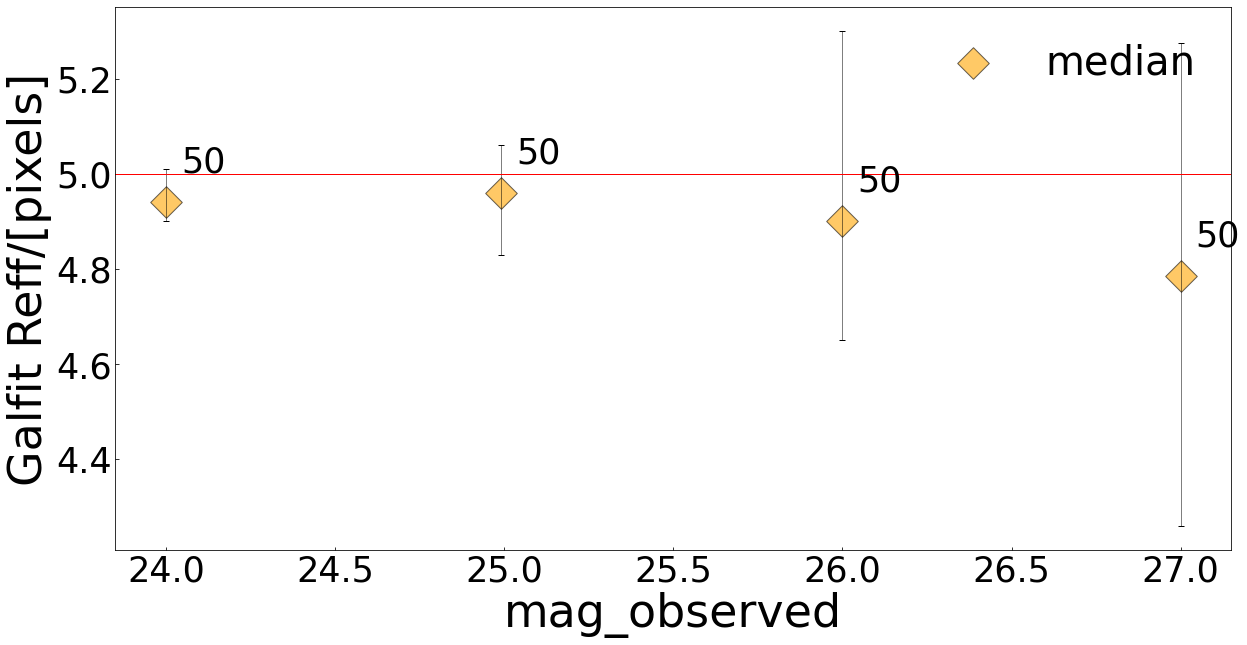}}
    \vspace{0.1cm}
    \subfloat{\includegraphics[ width=10cm, height=5cm]{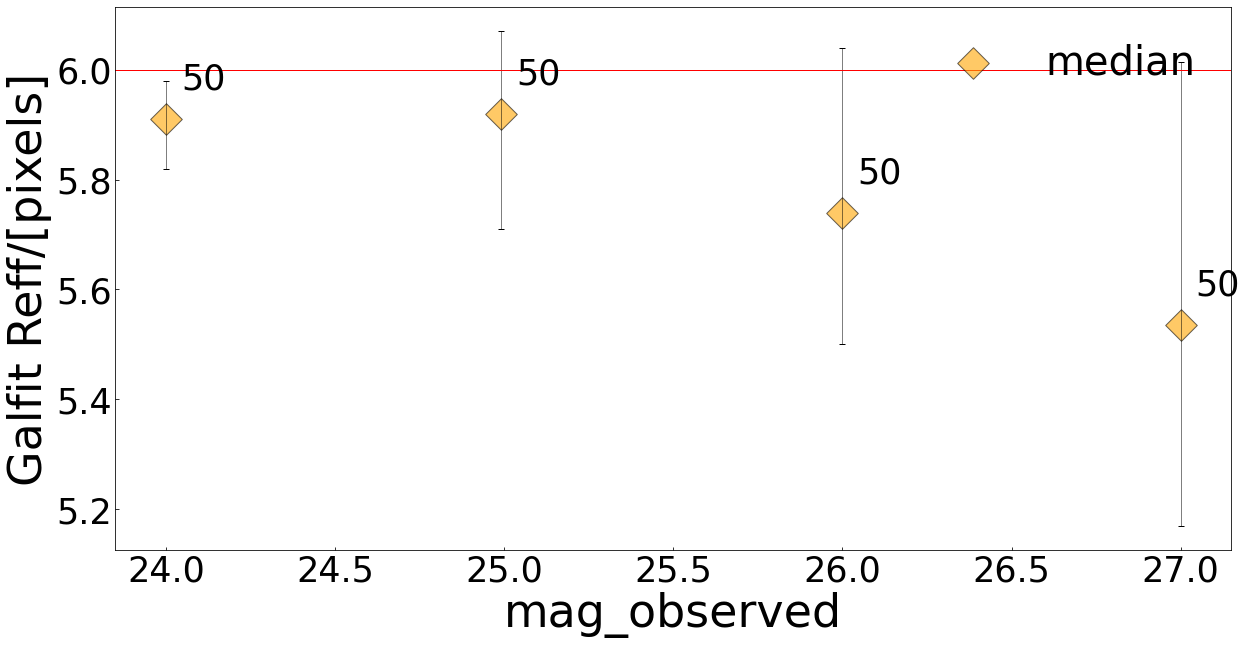}}
    \vspace{0.1cm}
    \subfloat{\includegraphics[ width=10cm, height=5cm]{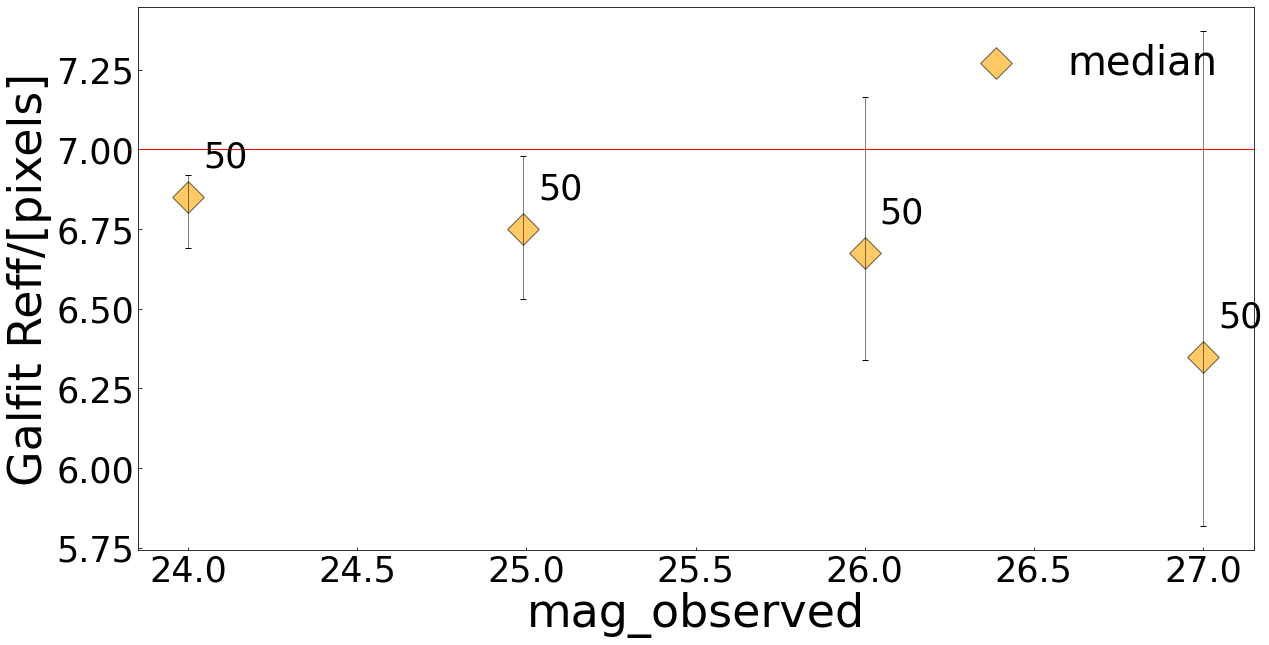}}

\caption{Same as Figure \ref{D1}. }
\label{D2}
\end{center}
\end{figure*}

Figures \ref{D1} and \ref{D2} summarize results from {\tt GALFIT} runs on a lensed mock images created by FM where we are testing accuracy of the size measurements.
\label{Galfit test}

\bsp	
\label{lastpage}
\end{document}